\documentclass{article}

\usepackage{graphicx}
\usepackage{balance}  

\usepackage{balance}
\usepackage{graphicx, comment}
\usepackage{cite}
\usepackage{url}

\usepackage{breakurl}
\usepackage{booktabs} 
\usepackage{tabularx}
\usepackage{algorithm, algorithmicx, algpseudocode}
\usepackage{pifont}
\usepackage{xspace}
\usepackage{caption}
\usepackage{subcaption}

\newcommand{\sysname}{Aegis\xspace}
\usepackage{arxiv}

\usepackage[utf8]{inputenc} 
\usepackage[T1]{fontenc}    
\usepackage{hyperref}       
\usepackage{url}            
\usepackage{booktabs}       
\usepackage{amsfonts}       
\usepackage{nicefrac}       
\usepackage{microtype}      
\usepackage{lipsum}

\title{Multifaceted Privacy: How to Express Your Online Persona
without Revealing Your Sensitive Attributes}

\author{
  Victor Zakhary \\
  Department of Computer Science\\
  UC Santa Barbara\\
  Santa Barbara, CA 93106 \\
  \texttt{victorzakhary@ucsb.edu} \\
   \And
  Ishani Gupta \\
  Department of Computer Science\\
  UC Santa Barbara\\
  Santa Barbara, CA 93106 \\
  \texttt{ishani@ucsb.edu} \\
  \And
  Rey Tang \\
  Department of Computer Science\\
  UC Santa Barbara\\
  Santa Barbara, CA 93106 \\
  \texttt{changreytang@ucsb.edu} \\
   \And
  Amr El Abbadi \\
  Department of Computer Science\\
  UC Santa Barbara\\
  Santa Barbara, CA 93106 \\
  \texttt{elabbadi@ucsb.edu} \\
}

\begin{document}
\maketitle

\begin{abstract}
Recent works in social network stream analysis show that a user's 
\textit{online persona} attributes (e.g., gender, ethnicity, 
political interest, location, etc.) can be accurately 
inferred from the topics the user writes about or engages
with. Attribute and preference inferences have been widely used to
serve personalized recommendations, directed ads, and to enhance the user
experience in social networks. However, revealing a user's sensitive attributes
could represent a privacy threat to some individuals. Microtargeting (e.g., \textit{Cambridge
Analytica scandal}), surveillance, and discriminating ads are examples of threats to user 
privacy caused by sensitive attribute inference. In this paper, we propose
{\em Multifaceted privacy}, a novel privacy model that aims to obfuscate a 
user's sensitive attributes while publicly preserving the user's public persona.
To achieve multifaceted privacy, we build \textit{\sysname}, a prototype client-centric social network stream processing system that helps 
preserve multifaceted privacy, and thus allowing social network users to 
freely express their online personas without revealing their sensitive 
attributes of choice. \sysname allows social network users to control which
persona attributes should be publicly revealed and which ones should be kept
private. For this, \sysname continuously suggests \textit{topics} and 
\textit{hashtags} to social network users to post in order to obfuscate 
their sensitive attributes and hence confuse content-based sensitive attribute
inferences. The suggested topics are carefully chosen 
to preserve the user's publicly revealed persona attributes while hiding their 
private sensitive persona attributes.
Our experiments show that adding as few as 0 to 4 obfuscation posts (depending
on how revealing the original post is) successfully hides 
the user specified sensitive attributes 
without changing
the user's public persona attributes
\end{abstract}

\keywords{Attribute Privacy, Online-Persona, Content-Based Inference}

\maketitle

\section{Introduction}\label{sec:introduction}
Over the past decade, social network platforms such as Facebook, Twitter, 
and Instagram have attracted hundreds of millions of 
users~\cite{fbInfo,twitterInfo, InstagramInfo}.  These platforms are   
widely and pervasively used to communicate, create online 
communities~\cite{georgiou2017extracting}, and socialize. Social media users 
develop, over time, online persona~\cite{zakhary2017locborg} that reflect their 
overall interests, activism, and diverse orientations.
Users have numerous
followers that are specifically interested in their personas and their postings which 
are aligned with these personas. However, due to the rise of machine 
learning and deep learning techniques, user posts and social network 
interactions can be used to accurately and automatically infer many user persona
attributes such as gender, ethnicity, age, political interest, and 
location~\cite{kosinski2013private,paul2011you,zheleva2009join, zhang2018tagvisor}. 
Recent work shows that it is possible to predict an individual user's 
location solely using content-based 
analysis of the user's posts~\cite{cheng2010you,chandra2011estimating}. Zhang et 
al.~\cite{zhang2018tagvisor} show that hashtags in user posts can alone be
used to precisely infer a user's location with accuracy of 70\% to 76\%. Also, Facebook 
likes analysis was successfully used to distinguish between Democrats and Republicans with 
85\% accuracy~\cite{kosinski2013private}.

Social network giants have widely used attribute inference to serve personalized
trending topics, to suggest pages to like and accounts to follow, and to notify 
users about hyper-local events. In addition, social networks such as Facebook use
tracking~\cite{fbpixel} and inference techniques to classify users into categories 
(e.g. Expats, Away from hometown, Politically Liberal, etc.). These categories are
used by advertisers and small businesses to enhance directed advertising campaigns.
However, recent news about the \textit{Cambridge Analytica scandal}~\cite{cambridgeBI} 
and similar \textit{data breaches}~\cite{otherBreachesCNBC} suggest that users cannot 
depend on the social network providers to preserve their privacy. User 
sensitive attributes such as \textit{\textbf{gender, ethnicity, and location}} have been 
widely misused in \textit{illegally 
discriminating ads}, \textit{microtargeting}, and \textit{surveillance}. A recent ACLU 
report~\cite{job_discrimination} shows that Facebook illegally allowed employers 
to exclude women from receiving their job ads on Facebook.
Also, several reports have shown that Facebook allows discrimination against some 
ethnic groups in housing ads~\cite{housing_discrimination}. News about the Russian-linked
Facebook Ads during the 2016 election suggests that the campaign targeted voters
in swing states~\cite{swingingfortune} and specifically in Michigan and 
Wisconsin~\cite{russiancnn}. In addition, location data collected from Facebook,
Twitter, and Instagram has been used to target activists of color~\cite{aclugeofeedia}.

An \textbf{online-persona} can be thought of as the set of user 
attributes that can be inferred about a user from their online 
postings and interactions. These attributes fall into two 
categories: \textit{public} and \textit{private} persona 
attributes. Users should \textbf{decide}
which attributes fall in each category. Some attributes (e.g., 
political orientation and ethnicity) should be publicly 
revealed as a user's followers might follow her because of her 
public persona attributes. Other attributes (e.g., gender and 
location) are private and sensitive, and the user would not like
them to be revealed. However, with the above mentioned inference methods, 
the social media providers, as well as any adversary receiving the user 
posting can reveal a user's sensitive attributes.

To remedy this situation, \textbf{in this paper}, we propose {\em multifaceted 
privacy}, a novel privacy model that aims to obfuscate a user's sensitive 
attributes while revealing the user's public persona attributes. Multifaceted 
privacy allows users to freely express their online public personas without 
revealing any sensitive attributes \textbf{of their choice}. For example, a 
\textit{\#BlackLivesMatter} activist might want to hide her 
location from the police and from discriminating advertisers while continuing to
post about topics specifically related to her political movement.
This activist can try to hide her location by disabling the geo-tagging feature
of her posts and hiding her IP address using an IP obfuscation browser
like Tor~\cite{torBrowser}. However, recent works have shown that 
content-based location inferences can successfully and accurately predict a user's 
location solely based on the content of her
posts~\cite{cheng2010you,chandra2011estimating,zhang2018tagvisor}. 
If this activist frequently posts about topics that discuss 
BLM events in Montpelier, Vermont, she is most probably a resident of Montpelier 
(Montpelier has a low African American population).

\begin{figure}[!htbp]
  \centering
  \includegraphics[width=0.5\columnwidth]{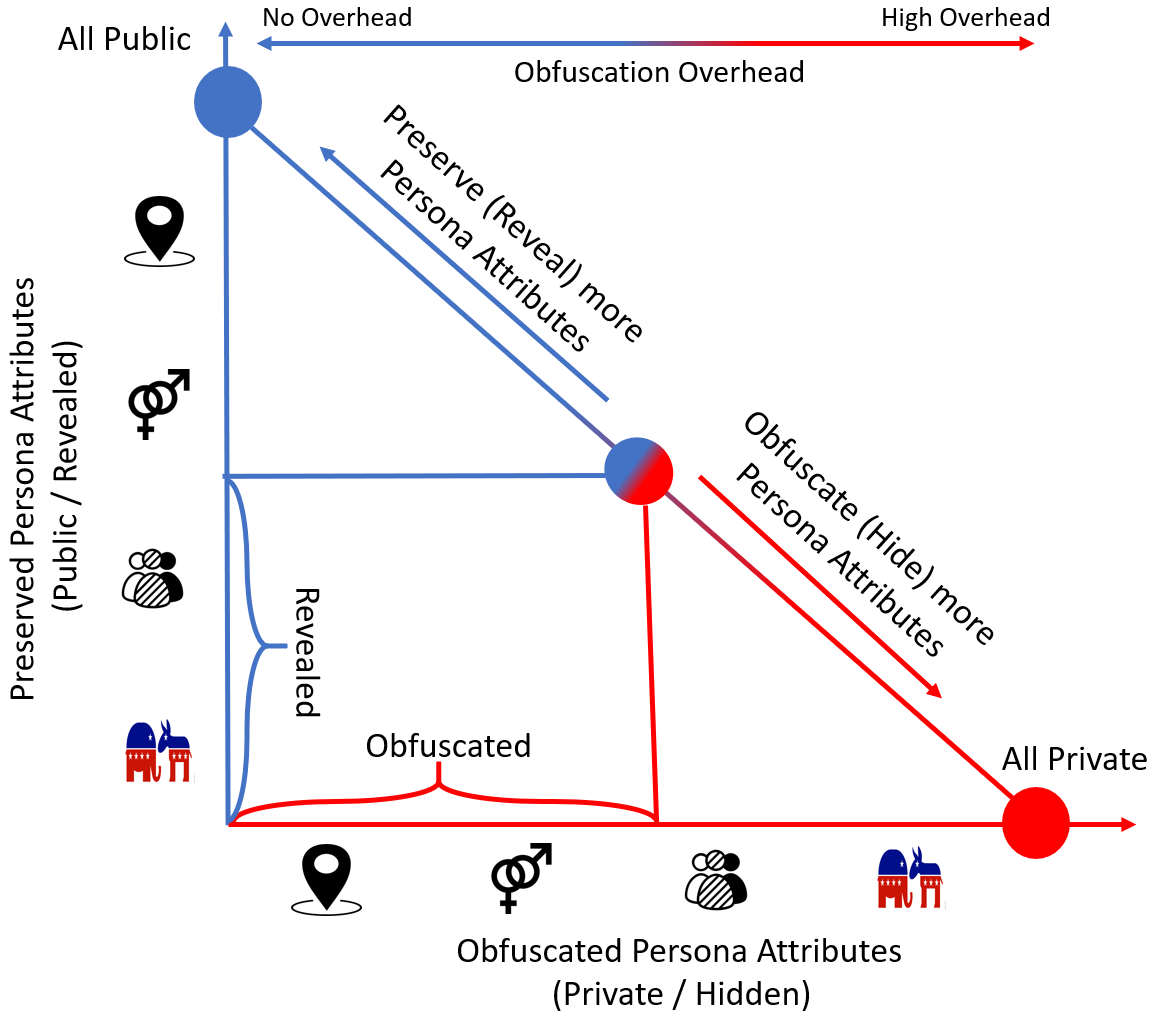}
  \caption{The dichotomy of multifaceted privacy: persona vs. privacy.}
  \label{fig:dichotomy}
\end{figure}

To achieve multifaceted privacy, we build \textit{\sysname}\footnote{\sysname: a shield in the Greek 
mythology.}, a prototype 
client-centric social network stream processing system that enables social 
network users to take charge of protecting their own privacy, instead of 
depending on the social network providers. Our philosophy in building 
\sysname is that social network users need to introduce some noisy interactions and obfuscation posts to
confuse \textbf{content based attribute inferences}. This idea is inspired from 
Rivest's chaffing and winnowing privacy model in~\cite{rivest1998chaffing}. 
Unlike in~\cite{rivest1998chaffing} where the sender and receiver exchange a secret 
that allows the receiver to easily distinguish the chaff from the wheat, in 
social networks, a user (sender) posts to an open world of followers (receivers)
and it is not feasible to exchange a secret with every recipient. In addition,
a subset of the recipients could be adversaries who try to infer the user's
sensitive attributes from their postings. 
Choosing this noise introduces a challenging \textit{dichotomy} and tension between the 
utility of the user persona and her privacy. Similar notions of
dichotomy between sensitive and non sensitive personal attributes
have been explored in sociology and are referred to as contextual 
integrity~\cite{nissenbaum2004privacy}.
Obfuscation posts need to be carefully
chosen to achieve obfuscation of private attributes without damaging the user's
public persona. For example, a \#NoBanNoWall activist loses persona utility if she 
writes about \#BuildTheWall to hide her location. Multifaceted privacy represents a
continuum between privacy and persona utility. Figure~\ref{fig:dichotomy} captures
this continuum, where the \textit{x-axis} represents the persona attributes that need to be 
obfuscated or kept private while the \textit{y-axis} represents the persona attributes that 
should be publicly preserved or revealed. Figure~\ref{fig:dichotomy} shows that 
privacy is the \textit{reciprocal} of the persona utility. Any attribute that needs 
to be kept private cannot be preserved in the public persona. As illustrated, the 
more attributes are kept private, the more obfuscation overhead is needed to achieve 
their privacy. A user who chooses to publicly reveal all her persona attributes 
achieves no attribute privacy and hence requires no obfuscation posting overhead.

As illustrated in Figure~\ref{fig:dichotomy}, a user chooses a point on the diagonal 
line of the multifaceted privacy that determines which persona 
attributes should be publicly revealed (e.g., political interest, ethnicity, etc) 
and which ones should be kept private (e.g., location, gender, etc). Note that users can reorder the attributes on the axes of 
Figure~\ref{fig:dichotomy} in order to achieve their intended 
public/private attribute separation.
Unlike previous approaches that require users to change their posts 
and \textit{hashtags}~\cite{zhang2018tagvisor} to hide their sensitive attributes, 
\sysname allows users to publish their original posts without changing their content. 
Our experiments show that adding obfuscation posts successfully preserves multifaceted
privacy.
\sysname considers the added noise as the cost to pay for 
achieving multifaceted privacy. Therefore, \sysname targets users who are 
willing to write additional posts to hide their sensitive attributes.

Aegis is user-centric, as we believe that users need to take control of their
own privacy concerns and cannot depend on the social media providers.  This is challenging
as it requires direct user engagement and certain sacrifices.  However, we believe Aegis will
help better understand the complexity of privacy as well as the role for individual engagement
and responsibility. Aegis represents a first step
in the long path to better understand the tensions between user privacy, the utility of social media,
and  trust in public
social media providers.  This is an overdue discussion that needs to be discussed by the scientific community,
and we believe Aegis will facilitate the medium for this discussion.

Our contributions are summarized as follows:
\begin{itemize}
\item We propose {\em multifaceted privacy}, a novel privacy model that
represents a continuum between the privacy of sensitive private attributes 
and public persona attributes.
\item We build \sysname, a prototype user-centric social network stream processing system 
that preserves multifaceted privacy.
\sysname continuously 
analyzes social media streams to suggest topics to post that are aligned with 
the user's public persona but hide their sensitive attributes.
\item We conduct an extensive experimental study to show that \sysname can 
successfully achieve multifaceted privacy.
\end{itemize}

The rest of the paper is organized as follows.  We explain the models of user, topic, and security in Section~\ref{sec:privacy_model}. Topic
classification algorithms and data structures that achieve multifaceted privacy are described 
in Section~\ref{sec:tree_structure} and 
\sysname's system design is explained in Section~\ref{sec:system_design}. Afterwards, an 
experimental evaluation is conducted in Section~\ref{sec:evaluation} to evaluate the  
effectiveness of \sysname in achieving the multifaceted privacy. 
The related work is presented in Section~\ref{sec:related_work} and future 
extensions are presented in Section~\ref{sec:future_extensions}. The paper 
is concluded in Section~\ref{sec:conclusion}.

\section{Models} \label{sec:privacy_model}

In this section, we present the user, topic, and security models.
The user and topic models explain how users and topics are represented in the system. 
The security model presents both the privacy and the adversary models. 

\subsection{User Model}\label{sub:user_model}
Our user model is similar to the user model presented in~\cite{georgiou2017privacy}.
The set $U$ is the set of social network users where $U = {\{u_1, u_2, ...\}}$.
A user $u_i$ is represented by a vector of attributes $V_{u_i}$  (e.g., gender $V_{u_i}[g]$, 
ethnicity $V_{u_i}[e]$, age $V_{u_i}[a]$, political interest $V_{u_i}[p]$, location $V_{u_i}[l]$, etc). Each 
attribute $a$ has a domain $a.d$ and the attribute values are picked from this domain. 
For example, the gender attribute $g$ has domain $g.d$ = \{male, female\}\footnote{Due to the limitation of the inference models, the gender attribute is considered only binary.
However, better models can be used to infer non binary gender attribute values.} and 
$\forall_{u_i \in U} V_{u_i}[g] \in g.d$. An example user $u_x$ is represented by 
the vector $V_{u_x}$ where $V_{u_x}$ = ($g$: female, $e$: African American, $a$: 23, $p$: 
Democrat, $l$: New York). Attribute domains can form a hierarchy (e.g., location: city 
$\rightarrow $ county $\rightarrow$ state $\rightarrow$ country) and an attribute 
can be generalized by climbing up this hierarchy. A user who lives in Los Angeles is also 
a resident of Orange County, California, and the United States. Other attributes can form
trivial hierarchies (e.g., gender: male or female $\rightarrow$ * (no knowledge)). 

The user attribute vector $V_{u_i}$ is divided into two main categories: 1) 
the set of public {\em persona} attributes 
$V_{{u_i}}^p$ and 2) the set of private {\em sensitive} attributes  $V_{{u_i}}^s$.
Multifaceted privacy aims to publicly reveal all persona attributes in $V_{{u_i}}^p$ while hiding 
all sensitive attributes in $V_{{u_i}}^s$. Each user defines her $V_{{u_i}}^s$ and $V_{{u_i}}^p$ {\em a priori}. 
As shown in Figure~\ref{fig:dichotomy}, attributes in $V_{{u_i}}^p$ are the complement of the
attributes in $V_{{u_i}}^s$. Therefore, each attribute either belongs to $V_{{u_i}}^p$ or $V_{{u_i}}^s$.

\subsection{Topic Model} \label{sub:topic-model}

The set $T$ represents the set of all topics that are discussed by all the
social network users in $U$. $T_i^\tau \subset T$ represents the set of all the topics 
posted by user $u_i$'s till time $\tau$ where
$T_i^\tau = \{t_i^1, t_i^2, ..., t_i^n\}$.
Social network topics are characterized 
by the attributes of the users who post about these topics. 
Unlike user attributes, which are discrete values, topic attributes are represented as distributions. 
For example, an analysis of the ethnicity of the users who 
post about the topic \texttt{\#BlackLivesMatter} can result in the distribution
10\% Asian, 25\% White, 15\% Hispanic, and 50\% Black. This distribution means 
that Asians, Whites, Hispanics, and Blacks post about the topic \#BlackLivesMatter and
50\% of the users who post about this topic are Black. A topic $t_i$ is represented by 
a vector of attribute distributions $V_{t_i}$ where $V_{t_i}[g]$, $V_{t_i}[e]$,
$V_{t_i}[p]$, and $V_{t_i}[l]$ are respectively the gender, the ethnicity, the political 
party, and the location distributions of the users who post about $t_i$.
 
To extract the gender, ethnicity, and political interest attribute
distributions of different topics, we use the language models introduced 
in~\cite{georgiou2017extracting, schwartz2013personality}.
However, any other available model could be used to infer
user attributes.
The location
distribution of a topic is inferred using the geo-tagged public posts about 
this topic, where the location of the publisher is explicitly attached to the post.

\subsection{Security Model} \label{sub:privacy_model}

An approach that is commonly used for attribute obfuscation is
{\em generalization}. The idea behind attribute generalization is to report a 
generalized value of a user's sensitive attribute in order to hide the actual
attribute value within. Consider location as a sensitive 
attribute example. 
Many works~\cite{mokbel2006new,zhang2018tagvisor} have used location 
generalization in different contexts. Mokbel et al.~\cite{mokbel2006new} use location 
generalization to hide a user's exact location from Location Based Services (LBS). A query
that asks \textit{"what is the nearest gas station to my exact location in Stanford, CA?"}
should be altered to \textit{"list all gas stations in California".} 
Notice that the returned 
result of the altered query has to be filtered at the client side to find the answer of the 
original query.
Similarly, Andres et al.~\cite{Andres:2013} 
propose geo-indistinguishability, a location privacy model that uses differential privacy to hide
a user's exact location in a circle of radius $r$ from
LBS providers. The wider the generalization range, the more privacy achieved, and the more network 
and processing overhead are added at the client side.
Similarly, in the context of social networks, Zhang et al.~\cite{zhang2018tagvisor} require
Twitter users to generalize their location revealing hashtags in order to hide their exact location.
For example, a user whose post includes "\#WillisTower" should be generalized to "\#Chicago" to hide
a user's exact location. Notice that generalization requires users to alter 
their original posts or queries.

To overcome these limitations and to allow users to write their posts using their own words, we adopt 
the notion of \textit{k-attribute-indistinguishability} privacy that is defined as follows. 
For every sensitive attribute $s \in V_{{u_i}}^s$, the user defines an indistinguishability parameter 
$k_s$. $k_s$ determines the number of attribute values among which the real value
of parameter $s$ is hidden. For example, a user who lives in CA sets $k_l = 3$ in order to hide
her original state location, CA, among 3 different states (e.g., CA, IL, and NY). This means that
a content-based inference attack should not be able to distinguish the user's real location
among the set \{CA, IL, NY\}. As explained in~\ref{sub:user_model}, attribute domains either
form multi-level hierarchies (e.g., location) or trivial hierarchies (e.g., gender and ethnicity). 
Unlike in attribute generalization where a user's attribute value is generalized by climbing up the
attribute hierarchy, k-attribute-indistinguishability achieves the privacy of an attribute value
by hiding it among $k-1$ attribute values chosen from the siblings of the actual attribute value in 
the same hierarchy level (e.g., a user's state level location is hidden among $k-1$ other states 
instead of generalizing it to the entire country). The following inference attack explains when
k-attribute-indistinguishability is achieved or violated.

\textbf{The adversary assumptions: } the adversary model and the inference attacks are 
similar to the ones presented in~\cite{zhang2018tagvisor}. However, unlike 
in~\cite{zhang2018tagvisor}, our inference attack is not only limit to the location 
attribute but also can be extended to infer every user sensitive attribute in $V_{{u_i}}^s$.
The adversary has access to the set of all topics $T$ and all the public 
posts related to each topic. This assumption covers any adversary who can 
crawl or get access to the public posts of every topic in the social network. 
As proposed in Section~\ref{sec:introduction}, the target user $u_i$ 
does not reveal her sensitive attribute values to the public (e.g., a user who wants to hide
her location must obfuscates her IP address and disable the geo-tagging feature for her
posts). Therefore, the adversary can only see the content of the public 
posts published by $u_i$. The adversary uses their knowledge about the set of all topics
$T$ and the set of topics $T_i^\tau$ discussed by $u_i$ to infer her 
sensitive attributes. Multifaceted privacy protects users against an adversary who only performs
\textit{content-based inference} attacks. Therefore, multifaceted privacy assumes that 
the adversary does not have any \textit{side channel knowledge} that can be 
used to reveal a user's sensitive attribute value (e.g., another online profile 
that is directly linked to the user $u_i$ where sensitive attributes such as gender, 
ethnicity, or location are revealed). 

\textbf{Inference attack:} the adversary's ultimate goal is to reveal or at least have
high confidence in the knowledge of the sensitive attribute values of the target user $u_i$. 
For this,
the adversary runs a content-based attack as follows. First, the adversary crawls the 
set of topics $T_i^\tau$ that user $u_i$ wrote about. For each topic, the adversary 
infers the demographics of the users who wrote about this topic. Then, the adversary 
aggregates the demographics of all the topics in $T_i^\tau$. The adversary uses the
aggregated demographics to estimate the sensitive attributes of user $u_i$.
The details of the inference attack is explained as follows.

For each topic $t_j \in T_i^\tau$, an adversary crawls the set of posts $P_{t_j}$
that discusses topic $t_j$ and for each post $p_i \in P_{t_j}$, the adversary 
uses some models to infer the gender, ethnicity, political interest, 
and location of the user who wrote this post. Then, the adversary uses
the inferred attributes of each post in topic $t_j$ to populate $t_j$'s 
distribution vector $V_{t_j}$. For example, $V_{t_j}[g]$ is the gender 
distribution of all users who wrote about topic $t_j$. Similarly, $V_{t_j}[e]$,
$V_{t_j}[p]$, and $V_{t_j}[l]$ are the ethnicity, political interest, and location
distributions of the users who posted about topic $t_j$. We define $V_{u_i}^*$ as a 
vector of attribute distributions that is used to estimate the attributes of user $u_i$.
$V_{u_i}^*$ is the result of aggregating the normalized $V_{t_j}$ for every topic
$t_j \in T_i^\tau$ as shown in Equation~\ref{equation:user_probability}.

\begin{equation}
\label{equation:user_probability}
V_{u_i}^*=  \frac{\sum_{t_j \in T_i^\tau} \frac{ V_{t_j}}{|P_{t_j}|}}{|T_i^\tau|}
\end{equation}

Equation~\ref{equation:user_probability} shows that the topic's attribute
distribution vector $V_{t_j}$ is first normalized by dividing $V_{t_j}$ by 
the number of posts in topic $t_j$. This normalization equalizes the effect of 
every topic $t_j \in T_i^\tau$ on the user's attribute estimations in $V_{u_i}^*$. 
$V_{u_i}^*$ is the summation of the normalized $V_{t_j}$ for every topic $ t_j \in 
T_i^\tau$ divided by the number of topics in $T_i^\tau$. $V_{u_i}^*[a]$ is the 
distribution of attribute $a$ for user $u_i$. For example, a user might have a gender
distribution $V_{u_i}^*[g]$ = \{female:0.8, male:0.2\}. This means that the inference
attack using $u_i$'s posted topics suggests that the probability $u_i$ is a female is 80\% while $u_i$ is a male is only 20\%. For 
every attribute $a$, an attacker uses the maximum attribute value $max(V_{u_i}^*[a])$ 
as an estimation of the actual value $V_{u_i}[a]$. In the previous example, an 
attacker would infer $V_{u_i}^*[g] = $ female
as an estimate of the gender of user $u_i$. An inference attack succeeds
in estimating an attribute $a$ if the attacker can have sufficient confidence in
estimating the actual value $V_{u_i}[a]$.  This confidence is achieved if the
difference between the maximum estimated attribute value of $a$ and the top-$k^{th}$
estimated attribute value of $a$ is greater 
than $\Delta_a$.
For example, if $\Delta_g = 0.1$ and $k_g = 2$ (assuming gender is
a binary attribute and it needs to be hidden among the 2 gender attribute values), 
then an attacker successfully estimates $u_i$'s gender if the 
$max(V_{u_i}^*[g])$ is distinguishable from the $2^{nd}$ highest value in $V_{u_i}^*[g]$
by more than 10\%. In the previous example where $V_{u_i}^*[g]$ = \{female:0.8, male:0.2\}, 
an attacker succeeds to estimate $u_i$'s gender = female as
the difference between $V_{u_i}^*[g].female - V_{u_i}^*[g].male \ge \Delta_g$.

\begin{figure}[ht!]
\centering
    \includegraphics[width=0.7\columnwidth]{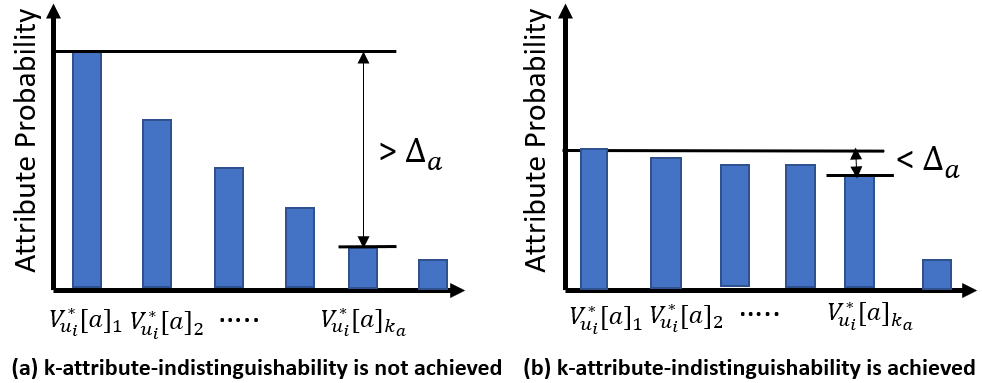}
    \caption{Illustrating k-attribute-indistinguishability }
    \label{sub:fig:privacy_model}
\end{figure}

Figure~\ref{sub:fig:privacy_model} show an example of a successful inference
attack and another of a failed inference attack on attribute $a$ of user $u_i$. As shown
in Figure~\ref{sub:fig:privacy_model}.a, the maximum estimated attribute
value $V_{u_i}^*[a]_1$ is distinguishable from the top-$k^{th}$ ($k_a$ for attribute
$a$) attribute values in $V_{u_i}^*[a]$ by more than $\Delta_a$. In this 
scenario, an attacker can conclude with high confidence that $V_{u_i}^*[a]_1$ is a 
good estimate for $V_{u_i}[a]$. However, in Figure~\ref{sub:fig:privacy_model}.b, 
$V_{u_i}^*[a]_1$ is indistinguishable from the top-$k^{th}$ attribute values in 
$V_{u_i}^*[a]$. In this scenario, the attack is marked failed and
k-attribute-indistinguishability is achieved.

The parameter $k$ is used to determine the number of attribute values within which
the user's actual attribute value is hidden. The bigger the $k$, the less the
attacker's confidence about the user's actual attribute value. 
As a result, increasing
$k$ introduces uncertainty in the attacker's inference and hence boosts the 
adversary cost to micro-target users who hide their actual
attribute values among $k$ different attribute values. For example, an
adversary who wants to target a user in location CA has to pay 3 times the 
advertisement cost to reach the same user if the user equally hides her 
location among 3 other locations (e.g., CA, IL, and NY).

We understand that the requirement to determine the sensitive attributes and
an indistinguishability parameter value $k_s$ for every sensitive attribute 
$s \in V_{{u_i}}^s$ could be challenging for many users. Users might not 
have a sense of the number of attribute values to obfuscate the 
actual value of a particular attribute. One possible solution to address this usability 
challenge is to design a questionnaire for \sysname's first time users. This
questionnaire helps \sysname understand which persona attributes
are sensitive and how critical the privacy of every
sensitive attribute is to every user. This allows \sysname
to auto-configure $k_s$ of every sensitive attribute $s$ for every user. 
The details
of such an approach is out of the scope of this paper. In this paper,
we assume that \sysname is preconfigured with the set $V_{{u_i}}^s$ and
for every attribute $s \in V_{{u_i}}^s$, the value of $k_s$ is determined.

\section{Multifaceted Privacy} \label{sec:tree_structure}

Multifaceted privacy aims to obfuscate a user's sensitive attributes
for every attribute in $V_{{u_i}}^s$.
This has to be done while publicly revealing every attribute in the user's
public persona in $V_{{u_i}}^p$. Various approaches have been used to obfuscate specific sensitive attributes, in particular, \textit{Tagvisor}~\cite{zhang2018tagvisor} protects users against content-based inference attacks by requiring users to alter their posts by changing or replacing hashtags
that reveal their sensitive attributes, in their case location.  Our approach is different, as it is paramount to not only preserving
the privacy of the sensitive attributes, but also to preserve the on-line persona of the user, and hence reveal their public attributes.  
It is critical for a user to post their posting in their own words that reflect their persona.  
We therefore preserve multifaceted privacy by hiding a specific post among other obfuscation posts.  
Our approach needs to suggest posts that are aligned with
the user's public persona but linked to alternative attribute values of their sensitive attributes
in order to obfuscate them. This requires a topic classification model that simplifies
the process of suggesting obfuscation posts.  
For example, consider State level location as 
a sensitive attribute. To achieve k-location indistinguishability,
a user's exact State should be hidden among $k-1$ other States. 
This requires suggesting obfuscation postings about topics that are 
linked to these other $k-1$ States. Users in NY state can use topics 
that are mainly discussed in IL to obfuscate their location among
NY and IL. To discover such potential topics, all 
topics that are discussed on a social network need to be classified 
by the sensitive attributes that need to be obfuscated, State level location in
this example. A topic is linked to some State if the maximum estimated
State location of this topic, $max(V_{t_i}[l])$, is distinguishable from
other State location estimates in $V_{t_i}[l]$ by more than $\Delta_l$.
For example, if $\Delta_l = 10\%$, a topic that has a State location 
distribution of \{NY=0.6, IL=0.2, CA=0.1, Others=0.1\} is linked to NY
State while a topic that does not have a distinguishable State location 
inference by more than $\Delta_l$ is not linked to any State. In this 
section, we first explain a simple but incorrect topic classification 
model 
that successfully suggests obfuscation posts that hide a user's sensitive attributes but does not preserve her public persona. Then, we explain 
how to modify the topic classification model to suggest obfuscation 
posts that do achieve multifaceted privacy.

\begin{figure}[!htbp]
  \centering
  \includegraphics[width=0.5\columnwidth]{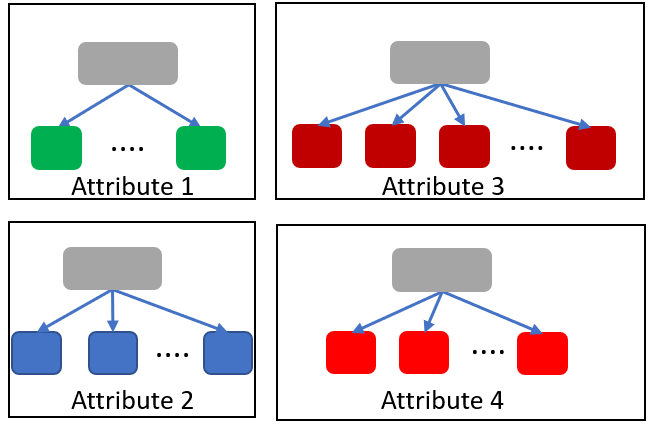}
  \caption{Each attribute independently forms a hierarchy} 
  \label{fig:reject_proposal}
\end{figure}

\textbf{A simple incorrect proposal: } in this proposal, topics are classified 
 \textbf{independently} by every sensitive attribute. As shown in 
 Figure~\ref{fig:reject_proposal}, each attribute forms an independent hierarchy. 
 The root of the hierarchy has the topics that are not linked to a specific 
 attribute value. Topics that are strongly linked to some attribute
 value fall down in the hierarchy node that represents this attribute
 value. For example, State level location attribute forms a hierarchy
of two levels. The first level, the root, has all the topics that do not belong
to a specific State. A topic like \#Trump is widely discussed in all the 
States and therefore it resides on the root of the location attribute. 
However, \#cowboy is mainly discussed in TX and therefore it falls down in 
the hierarchy to the TX node. To obfuscate a user's State location, topics need to 
be selected from the sibling nodes of the user's State in the State level location 
hierarchy. These topics belong to other locations and can be used to achieve 
k-location-indistinguishability privacy. Although this proposal
successfully achieves location privacy, the suggested posts are not necessarily 
aligned with the user's public online persona. For example, this obfuscation technique could
suggest the topic \#BuildTheWall (from TX) to an activist (from NY) who frequently posts
about \#NoBanNoWall in order to hide her location. This misalignment between the suggested
obfuscation posts and the user's public persona discourages users from seeking privacy
fearing the damage to their public online persona.

\begin{figure}[!htbp]
  \centering
  \includegraphics[width=0.8\columnwidth]{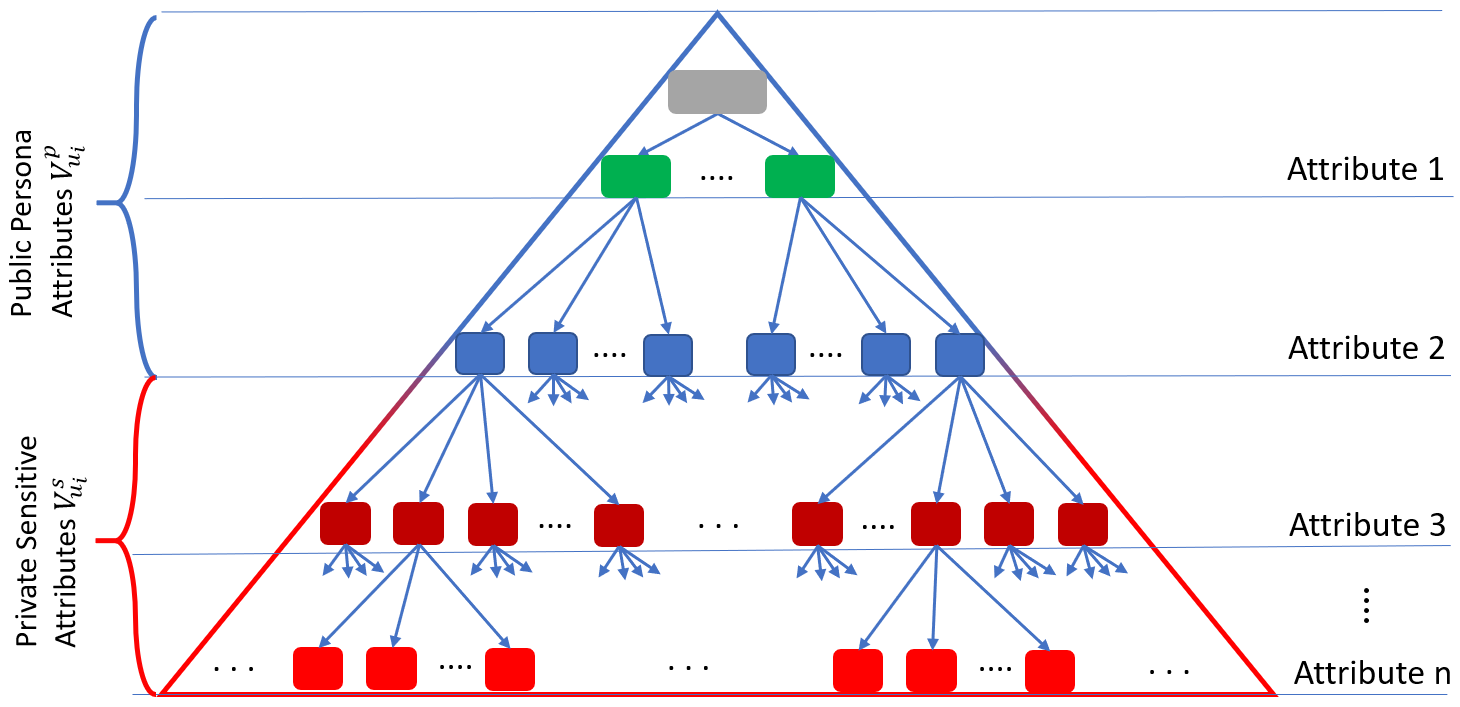}
  \caption{A dependent topic tree where public
  attributes are at the top while private attributes are at the bottom}
  \label{fig:tree_structure}
\end{figure}

\textbf{A persona preserving proposal: } To overcome the independent classification 
shortcomings, obfuscation postings need to be suggested from a tree hierarchy that 
captures relevant dependencies among all user attributes including both public
persona and private sensitive attributes. Public attributes need to reside on the upper
levels of the classification tree while private ones reside on the bottom
levels of the tree as shown in Figure~\ref{fig:tree_structure}. To achieve
k-attribute-indistinguishability, sibling values of the sensitive attributes
are used to hide the actual value of these sensitive attributes. By placing
the public attributes higher up on the tree, we ensure that the suggest topics
adhere to the public persona. Finally, multifaceted privacy only requires 
all the public attributes, regardless of their order, to reside on the 
upper levels of the tree while all the private attributes, regardless of 
their order, to reside on the lower levels of the tree. 

 Social network 
topics are \textbf{dependently} classified in the tree by the attribute domain values at each level.
For example, if the top most level of the tree is the political party attribute, topics 
that are mainly discussed by Democrats are placed in the left green child while topics that are mainly discussed by Republicans are place in the right green child. Note that topics 
that have no inference reside in the root of the hierarchy. Now, if the second public 
persona attribute is ethnicity (shown as blue nodes in Figure~\ref{fig:tree_structure}), 
topics in both the Democratic and Republican nodes are classified by the ethnicity domain 
attribute values. For example, topics that are mainly discussed by White Democrats are 
placed under the Democratic node in the White ethnicity node while topics that are mainly 
discussed by Asian Republicans are put under the Republican node in the Asian ethnicity 
node. This classification is applied at every tree level for every attribute. Now, assume a 
user is White, Female, Democrat, who lives in CA and wants to hide her location (shown
as red nodes in Figure~\ref{fig:tree_structure}) while publicly revealing her ethnicity, 
gender, and political party. In this case,
topics that reside in the sibling nodes of the leaf of her persona path, e.g., topics that are mainly discussed by White Female
Democrats who live in locations other than CA (e.g., NY and IL) can be suggested as obfuscation topics. This dependent 
classification \textbf{guarantees} that the suggested topics are aligned with the user's public persona but belong to other sensitive attribute values (different locations in this example). Note
that this technique is general enough to obfuscate any attribute and any number of
attributes. Each user defines her sensitive attribute(s) and the classification tree would be constructed with these attributes
to the bottom thus guaranteeing that the suggested posts do not
violate multifaceted privacy.

\section{\sysname System Design}\label{sec:system_design}


This section presents \sysname, a prototype social network stream 
processing system that implements 
multifaceted privacy and overcomes the adversarial content-based attribute 
inference attacks discussed in Section~\ref{sub:privacy_model}.
\sysname achieves k-attribute-indistinguishability by suggesting topics 
to post that are aligned with the social network user's public persona while
hiding the user's sensitive attributes among their other domain values. 
To achieve this, \sysname uses the classification and suggestion models 
discussed in Section~\ref{sec:tree_structure}. \sysname is designed in a 
\textit{user-centric}
manner which is configured on the user's local machine. In fact,
\sysname can be developed as a browser extension where all the user 
interactions with the social network are handled through this extension.
Every local deployment of \sysname only needs to construct a partition of the 
attribute-based topic classification hierarchy developed in 
Figure~\ref{fig:tree_structure}. This partition (or sub-hierarchy) 
include the user's attribute path from the root to a leaf in addition to the sibling
nodes of the user's sensitive attribute nodes. For example, a user whose 
attributes are Female, White, Democrat, and CA and wants to obfuscate 
her State location only requires \sysname to construct the Female, White, Democrat,
CA path in addition to $k-1$ other paths with the shared prefix Female, White,
Democrat but linked to $k-1$ other States. These $k-1$ States are used
to hide the user's true state in order to achieve k-location-indistinguishability. 
Note that if another user's public persona is specifically associated with their
location while they consider their ethnicity to be sensitive, then the hierarchy
needs to be reordered to reflect this criteria.

Although \sysname can be integrated with different online social network platforms, 
our prototype implementation of \sysname only supports Twitter to 
illustrate \sysname's functionality. Twitter provides developers with
several public APIs~\cite{twitterAPIs} that allow them to stream tweets 
that discuss certain topics. In addition, Twitter streaming APIs allow 
developers to sample 1\% of all the tweets posted on Twitter. In Twitter,
a topic is represented by either a \textit{hashtag} or a \textit{keyword}.
\sysname is built to work for \textit{new} Twitter profiles in order to 
continuously confuse an adversary about a profile's true sensitive attribute 
values from the genesis of this profile. \sysname
is not designed to work with existing old profiles as an adversary could have
already used their existing posts to reveal their sensitive attribute values. 
Even though \sysname suggests obfuscation posts, an adversary can distinguish 
the \textbf{old} original posts from the \textbf{newly} added original posts
accompanied by their obfuscation posts and hence reveals the user's sensitive
attributes true values.

\sysname is designed to achieve the following goals:

\begin{enumerate}
    \item to automate the process of streaming and classifying
    Twitter topics according to their attributes,
    \item to construct and continuously maintain the topic classification 
    sub-hierarchy,
    \item and finally to use the topic classification sub-hierarchy
    to suggest topics to the user that achieve multifaceted 
    privacy.
\end{enumerate}

To achieve these goals, \sysname consists of two main processes: 
\begin{itemize}
\item a \textbf{T}witter analyzer \textbf{P}rocess $TP$ and
\item a topic \textbf{S}uggestion \textbf{P}rocess $SP$.
\end{itemize}

\begin{figure}[!htbp]
  \centering
  \includegraphics[width=0.5\textwidth]{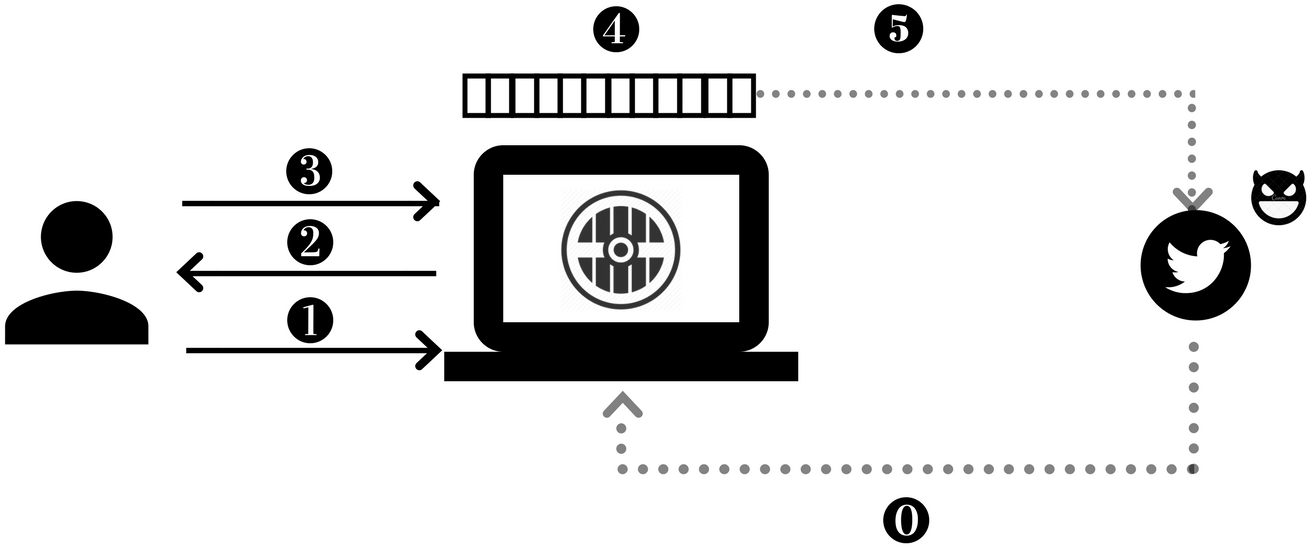}
  \caption{Aegis System Design and User Interaction Flow}
  \label{fig:system-design}
\end{figure}

$TP$ continuously analyzes the topics that are being discussed on 
Twitter and for each topic $t_i$, $TP$ uses the topic attribute inference 
models to infer $t_i$'s attribute distribution vector $V_{t_i}$. The accuracy of 
the topic attribute inference increases as the number of posts that discuss
topic $t_j$, $|P_{t_j}|$, increases. $TP$ uses a local key-value store as a topic 
repository where the key is the topic id $t_j \in T$ and the value is the 
topic attribute distribution vector $V_{t_i}$. In addition, $TP$ constructs and 
continuously maintains the topic classification sub-hierarchy that classifies 
the topics based on their attributes. This topic classification sub-hierarchy is 
used for suggesting obfuscation topics.
For this, the user provides $TP$ with both their attribute vector values  $V_{u_i}$ 
and their sensitive attribute vector $V_{u_i}^s$. $V_{u_i}$ and  $V_{u_i}^s$ 
determine the sub-hierarchy that $TP$ needs to maintain in order to obfuscate 
the attributes in $V_{u_i}^s$.
Figure~\ref{fig:system-design} shows the
interactions among the user, \sysname, and the social network.
As shown, step 0 represents the continuous Twitter stream 
analysis performed by $TP$. As $TP$ analyzes Twitter streams, it continuously
updates the topic repository and the classification sub-hierarchy.

The {\em topic suggestion process} $SP$ mainly handles user interactions with 
Twitter. $SP$ uses the topic classification sub-hierarchy constructed and
maintained by $TP$ to suggest obfuscation topics. For every sensitive attribute $s 
\in V_{u_i}^s$, the user provides the indistinguishability parameter $k_s$ that 
determines how many attribute values from the domain of $s$ should be used to hide
the true value of $s$. \sysname allows users to configure the privacy
parameter $\Delta_s$ for every attribute $s$. However, to enhance usability,
\sysname maintains a default value for the privacy parameter $\Delta_s = 10\%$. This means that the privacy of attribute $s$ is achieved if the 
inference attack cannot distinguish the maximum inferred attribute value from
the $k_s^{th}$ inferred attribute value by more than 10\%. 

$SP$ uses $k_s$, and $\Delta_s$ for every attribute $s \in V_{u_i}^s$ to 
generate the topic suggestion set $S_i$. Note that $SP$ is locally 
deployed at the user's machine. Therefore, the user does not have to trust 
any service outside of her machine. \sysname is designed to transfer user 
trust from the social network providers to the local machine. 
For every sensitive attribute $s$, $SP$ selects a fix set of 
$k_s-1$ attribute domain values. These $k_s-1$ attribute values are used to obfuscate the true value of attribute $s$, $V_{u_i}[s]$.

As shown in Figure~\ref{fig:system-design}, in Step 1, the user writes a post to 
publish on Twitter. $SP$ receives this post and queries $TP$ about the 
attributes of all the topics mentioned in this post. $SP$ uses these topic
attributes to simulate the adversarial attack. If $TP$'s topic inference
indicates that k-attribute-indistinguishability is violated for any attribute
$s \in V_{u_i}^s$, $SP$ queries $TP$ for
topics with public persona $V_{u_i}^p$ but linked to the other attribute
values of $s$ in the set of $k_s-1$ attribute values. For every returned topic, $SP$ 
ensures that writing about this topic enhances the aggregated inference
of the original post and the obfuscation posts towards 
$k_s$-attribute-indistinguishability. $SP$ adds these topics to the set $S_i$ and 
returns them to the user (Step 2). The user selects a few topics from $S_i$ to 
post in Step 3 and submits the posts to $SP$. Note
that users are \textbf{required to  write}  the  obfuscation posts using 
their personal writing styles to ensure that the original posts and the
obfuscation  posts  are indistinguishable~\cite{writingStyle2008,writingStyle2016}. 
Afterwards, $SP$ ensures that the
aggregated inference of submitted obfuscation posts in addition to the
original post lead to k-attribute-indistinguishability. 
Otherwise, $SP$ keeps suggesting more topics. As every original post 
along with its obfuscation posts achieve
k-attribute-indistinguishability, the aggregated inference over the 
whole user's posts achieve  
k-attribute-indistinguishability. In Step 4, $SP$ queues the original 
and the obfuscation posts and publishes them on the user's behalf in random
order and intervals 
to prevent \textit{timing attacks} 
(Step 5). An adversary can perform a timing attack if the original posts
and the obfuscation posts are distinguishable. Queuing and randomly 
publishing the posts prevents the adversary from distinguishing original 
posts from the obfuscation posts and hence prevents timing attacks.

We understand that the obfuscation writing overhead might alienate users 
from Aegis.  As a future extension, Aegis can exploit deep  neural  network 
language  models to learn the user’s writing style~\cite{ding2017learning}. Aegis can use such a 
model to generate~\cite{graves2013generating} full posts instead of hashtags and users can either 
directly publish these posts or edit them before publishing.

\section{Experimental Evaluation}\label{sec:evaluation}

In this section, we experimentally evaluate the effectiveness
of \sysname in achieving multifaceted privacy. 
We first present the experimental
setup and analyze some properties of the used dataset 
in Section~\ref{sub:experiment_setup}. Then, Sections~\ref{sub:inference_example} and ~\ref{sub:obfuscation_example} present illustrative inference and obfuscation examples 
that show the functionality of \sysname using real Twitter topics.
We experimentally show how \sysname can be used to hide
user location in Section~\ref{sub:location_obfuscation} and 
measure the effect of changing the indistinguishability parameter $k$ on 
the obfuscation overhead in Section~\ref{sub:k_effect}. Finally, Section~\ref{sub:gender_obfuscation}
illustrates the efficiency of \sysname on hiding the user 
gender while preserving other persona attributes. 

\subsection{Experimental Setup}\label{sub:experiment_setup}
\begin{figure*}[!htbp]
    \centering
    \begin{subfigure}[t]{0.32\textwidth}
        \includegraphics[width=\columnwidth]{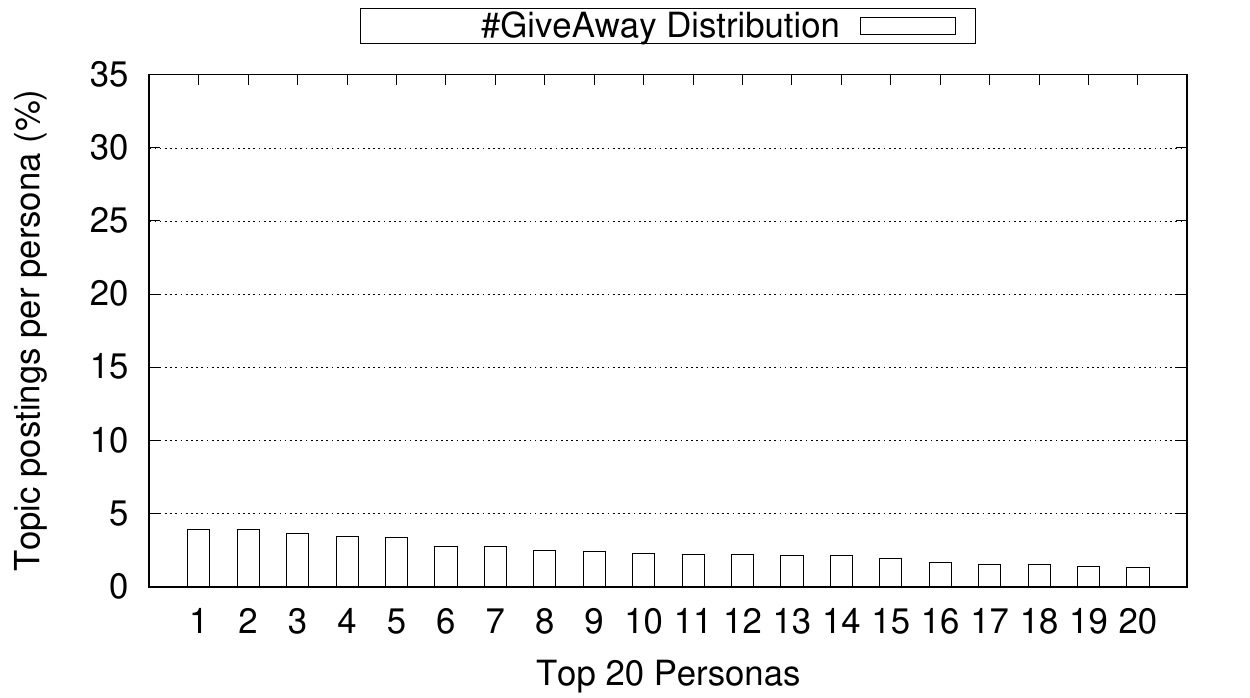}
        \caption{\#GiveAway (Negligible Connection)}
        \label{fig:GiveAway}
    \end{subfigure}
    \begin{subfigure}[t]{0.32\textwidth}
        \includegraphics[width=\columnwidth]{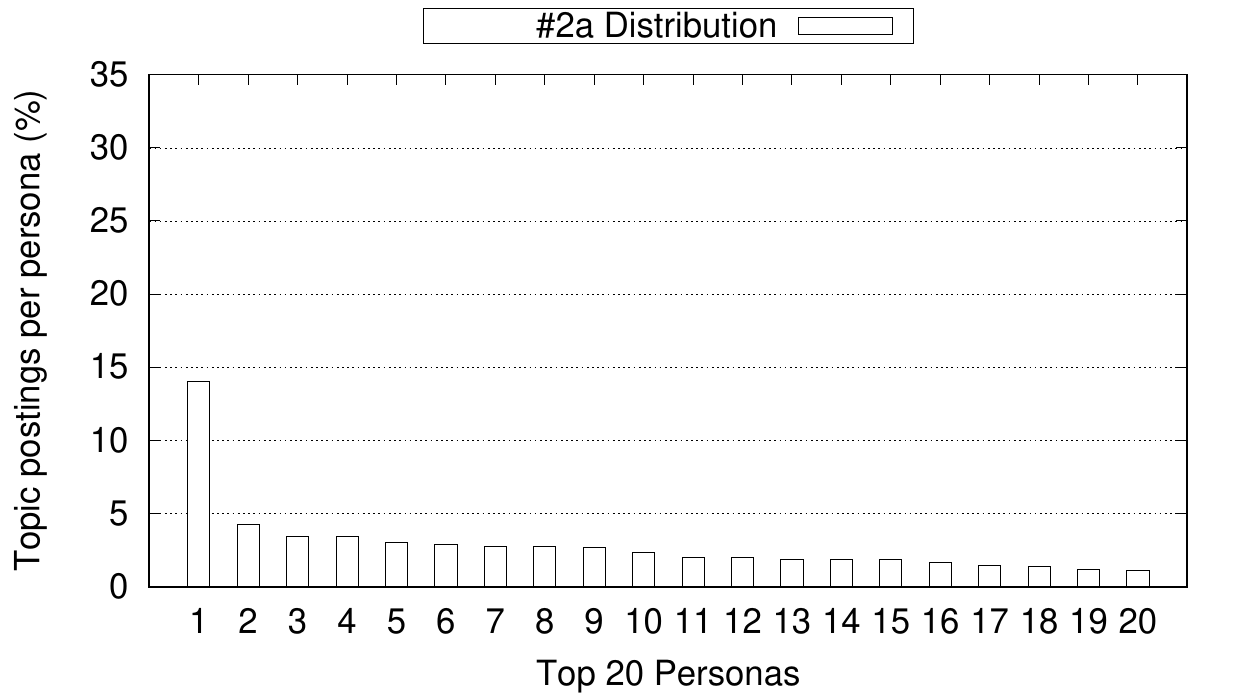}
        \caption{\#2a (Weak Connection)}
        \label{fig:2a}
    \end{subfigure}
    \begin{subfigure}[t]{0.32\textwidth}
        \includegraphics[width=\columnwidth]{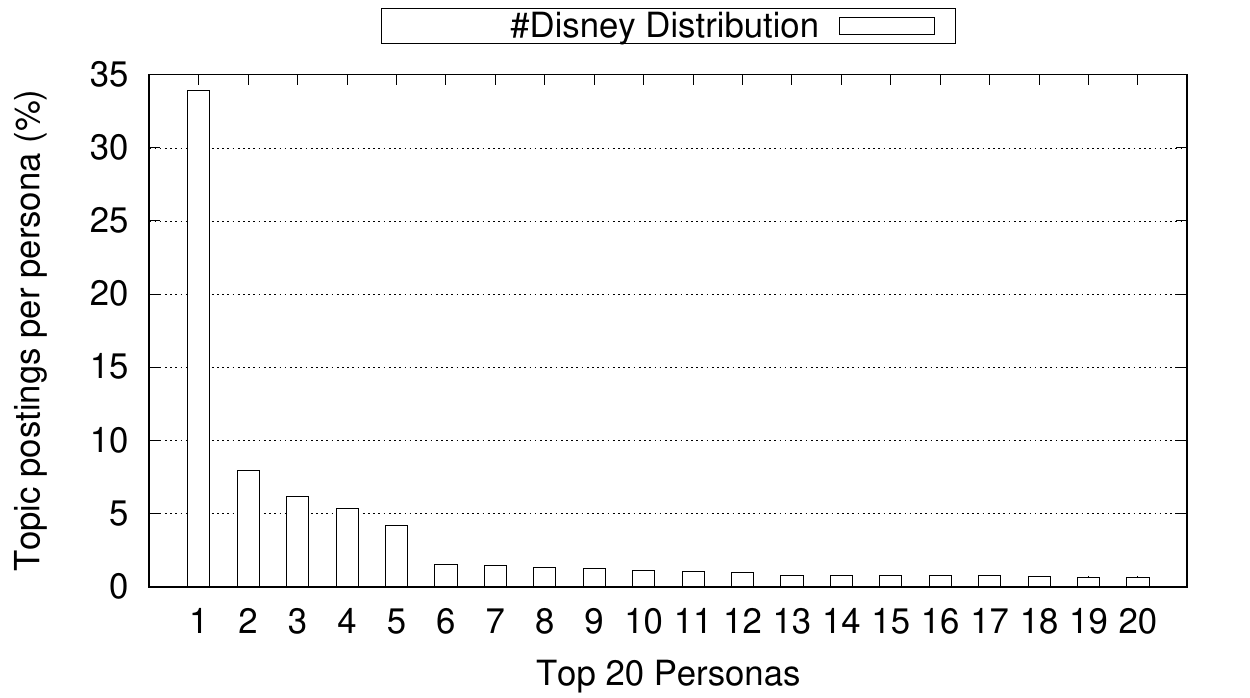}
        \caption{\#Disney (Strong Connection)}
        \label{fig:disney}
    \end{subfigure} 
    \caption{Examples of negligible, weak, and strong connection distribution topics for top-20 personas} 
    \label{fig:topic_connection}
\end{figure*}
Although \sysname can be integrated with different social network
platform, our prototype implementation is integrated
only with the Twitter social network. Twitter provides developers with
several public APIs~\cite{twitterAPIs} including the streaming
API that allows developers to crawl 1\% of all the postings on Twitter.
For our experiments, we use the 1\% random sampling of the Twitter stream 
during the year 2017. The attributes gender, ethnicity, and location are used to build a three
level topic classification hierarchy that classifies all topics according
to their attribute distribution.
For simplicity and without loss of generality,
we use the language models in~\cite{schwartz2013personality} to infer
both gender and ethnicity attributes of a post writer. In addition, we
infer the location distribution of different topics using the
explicitly geo-tagged posts about these topics. In the 1\% of Twitter's 
2017 postings, our models were able to extract 2,126,791 unique topics.
Our classification hierarchy suggests that 66\% of the dataset tweets are posted
by males. This analysis is consistent with the statistics published 
in~\cite{twitterGender}. In addition, the dataset has 6,864,300 
geo-tagged posts, 15\% of which originated in
California. Finally, the predominant ethnicity extracted from the dataset is White.

As the classification hierarchy is built using only gender, ethnicity, and state location
attributes, this results in a hierarchy of 500 different paths from the root to a leaf of
the hierarchy. These 500 paths result from all the possible combinations of
gender (male, female), ethnicity (White, Black, Asian, Hispanic, Native American), 
and state location (50 States). The 500 paths represent the different 500 personas
considered in our experiments. Our topic classification hierarchy suggests that topics
vary significantly in their connection to a specific persona path (a gender, ethnicity, and location
combination). For example, \#GiveAway is widely discussed among the 500 personas across the
50 States with very little skew towards specific personas over others. For topics that are widely 
discussed across all different persona, their skew is usually proportional to the population 
density of different States. For example, the top five highly populated states (CA, TX, FL, NY, 
and PA) usually appear as the top locations where widely discussed topics are posted.
On the other hand, other topics show strong connection to specific personas. For example,
33.93\% of the personas who write about \#Disney are Male, Asian, and live in Florida where Disney
World is located. Figure~\ref{fig:topic_connection} shows 3 examples of topics that
have trivial, weak, and strong connection to specific personas. In 
Figure~\ref{fig:topic_connection}, the x-axis represents the top-20 personas who
post about a topic and the y-axis represents the percentage of postings for each
persona. As shown in Figure~\ref{fig:GiveAway}, \#GiveAway has slight skew (negligible 
connection) towards some personas over other personas. Also, the top most persona who post
about \#GiveAway represent only 3.94\% of the overall postings about the topic. On the 
other hand, Figure~\ref{fig:2a} shows that the persona distribution for \#2a (refers to the
second amendment) has more skew (weak connection) towards some personas over others. As 
shown, the top posting persona on the topic \#2a contributes 14.03\% of the overall postings
of this topic. Finally, a topic like \#Disney has remarkable skew (strong connection)
towards some personas over others. As shown in Figure~\ref{fig:disney}, the top posting 
persona on the topic \#Disney contributes 33.93\% of the overall postings of this topic.

\begin{table}[!htbp]
 \centering
 \begin{tabular}{|l c c|} 
 \hline
 \textbf{Strength} & \textbf{Minimum $\delta$} & \textbf{Maximum $\delta$} \\  
 \hline
 Negligible & 0\% & 10\%  \\ \hline
 Weak & 10\% & 20\%  \\ \hline
 Mild & 20\% & 30\% \\ \hline
 Strong & 30\% & 100\% \\ \hline
 \end{tabular}
 \caption{Topic to persona connection strength categories and their corresponding $\delta$ ranges}
 \label{table:connection_strength}
\end{table}

\begin{table*}[ht!]
 \centering
 \begin{tabular}{|l c c c c c|} 
 \hline
 \textbf{Topic} & \textbf{Freq} & \textbf{M W Ca} & \textbf{F W Ca} & \textbf{M W Tx} & \textbf{F W Tx}\\  
 \hline
\#teen   &    7094 &    7   & \textbf{1}  &  15 &   4 \\ \hline  
\#hot    &  7478  &      5 &  \textbf{1} &   13  &   3 \\ \hline  
\#etsy   & 2739    &    6&   5  &  27  &  \textbf{1}  \\ \hline  
\#diy  &    1987  &      3  & 7 &   11 &    \textbf{1} \\ \hline
\#actor  & 725    &   \textbf{1}  &    2  &   9  & 11 \\ \hline
\#cowboys & 797  &   5  &    16  &  \textbf{1} &  2 \\ \hline
 \end{tabular}
 \caption{Topic Analysis By Persona}
 \label{table:topic_analysis}
\end{table*}
We define a {\em topic to persona connection strength} parameter $\delta$. $\delta$ is defined 
as the difference in posting percentage between the top-1 posting persona and
the top-k posting persona. For \#2a and k = 3, $\delta = 14.03 - 3.45 = 10.6$ while 
for \#Disney and for k=3, $\delta = 33.93 - 6.19 = 27.74$. We categorize topics into 4 
categories according to their $\delta$ value. As shown in Table~\ref{table:connection_strength}, a topic to persona connection that ranges
from 0\% to 10\% represents a {\em negligible connection} and hence this topic does not reveal
the persona attributes of the users who write about it. As the topic to persona
connection increases, the potential of revealing the attributes of the users who
write about this topic increases. In our experiments, we measure the overhead
of obfuscation for three different distributions with {\em weak, mild} and {\em strong} connections having a topic to persona strength connection ranges that are shown in Table~\ref{table:connection_strength}

\subsection{Illustrative Inference Example}\label{sub:inference_example}

Using our tree data structure we can infer interesting information
about different topics on Twitter in our dataset, specifically regarding
correlations between persona and topics. In Table~\ref{table:topic_analysis}
we analyze 7 topics and their connection to 4 of the most prominent personas in 
our dataset (Male-White-CA, Female-White-CA, Male-White-TX, Female-White-TX). {\em Frequency} denotes the number of times
the topic was observed and the number under a persona for a particular topic
denotes the order or rank in which a persona discusses this topic most.
For example, among all the persona we analyze in our dataset, \#actor is most discussed by
White Male Californians (rank 1) closely followed by White Female also from California
(rank 2). This is followed by other persona out of the focus of 
Table~\ref{table:topic_analysis}, until White Female Texans are reached at rank 9 and 
White Male Texans at rank 11. Also, Table~\ref{table:topic_analysis} shows that both
\#teen and \#hot are discussed the most by Female White Californians and 
that overall Females (in both CA and TX) who discuss this topic are more than Males in both CA and TX. We can also observe correlations across topics that have
semantic connections like \#etsy and \#diy. Etsy is an online platform
for users to sell DIY (Do It Yourself) projects. Female White Texans are 
much more interested in such DIY specific topics than any other of the personas.
Lastly, high correlations of certain topics can be observed with specific locations such 
as \#actor with California and \#cowboys with Texas. As Table~\ref{table:topic_analysis} 
reveals, the topics you post on social media significantly reveal your attributes, even 
private sensitive attributes you are unwilling to share. As such, we need a tool 
like \sysname to prevent adversaries from inferring private attributes while preserving
others.

\subsection{Illustrative Obfuscation Example}\label{sub:obfuscation_example}

\begin{figure}[!htbp]
  \centering
  \includegraphics[width=0.5\columnwidth]{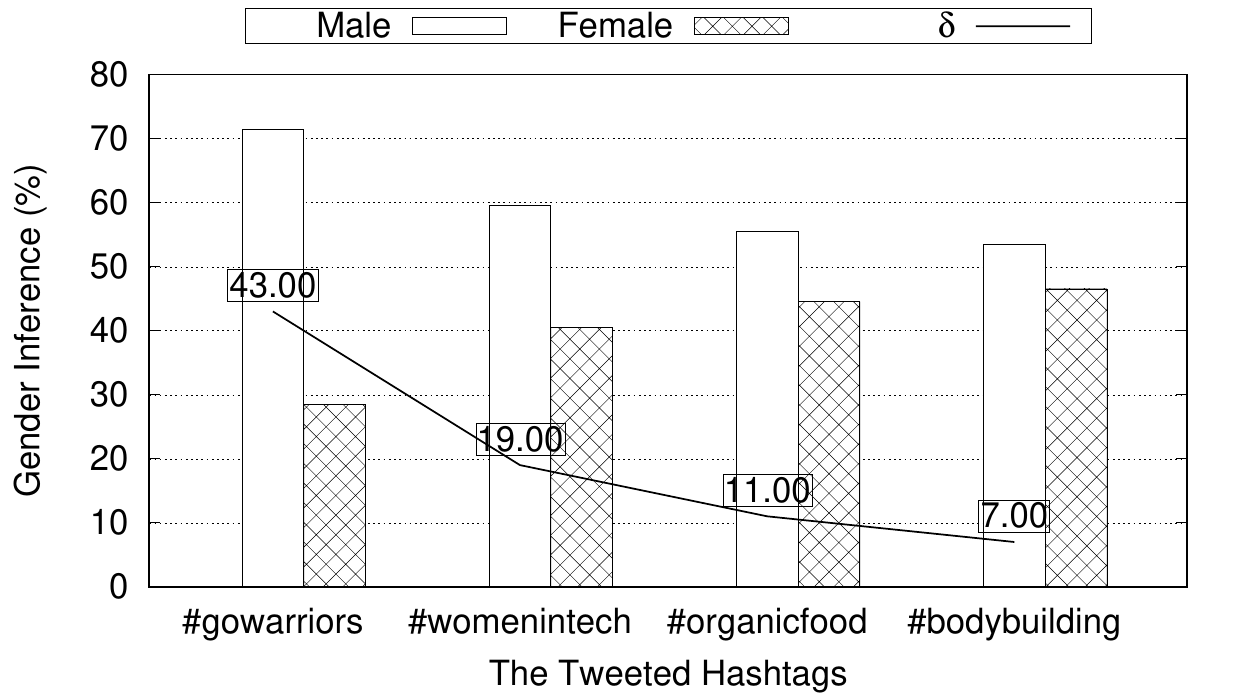}
  \caption{Illustrative Gender Obfuscation Example.}
  \label{fig:example-suggestions}
\end{figure}

\begin{figure*}[!htbp]
    \centering
    \begin{subfigure}[t]{0.32\textwidth}
        \includegraphics[width=\columnwidth]{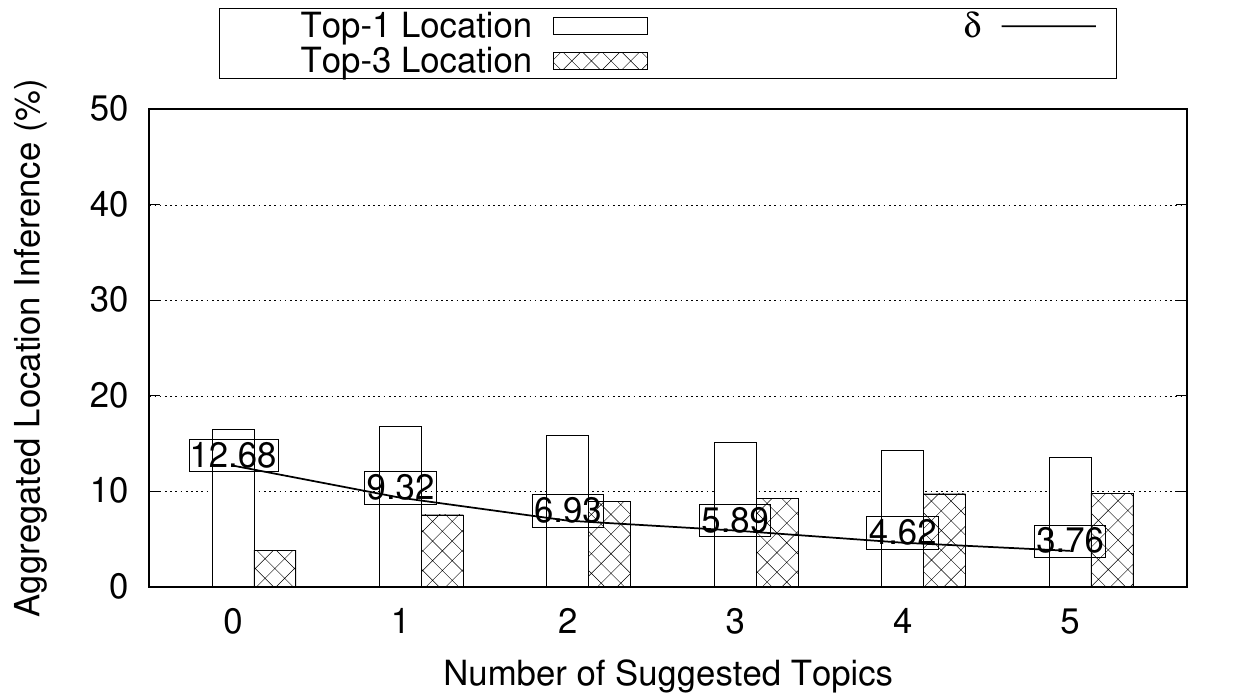}
        \caption{Location (Weak)}
        \label{fig:l_w}
    \end{subfigure}
    \begin{subfigure}[t]{0.32\textwidth}
        \includegraphics[width=\columnwidth]{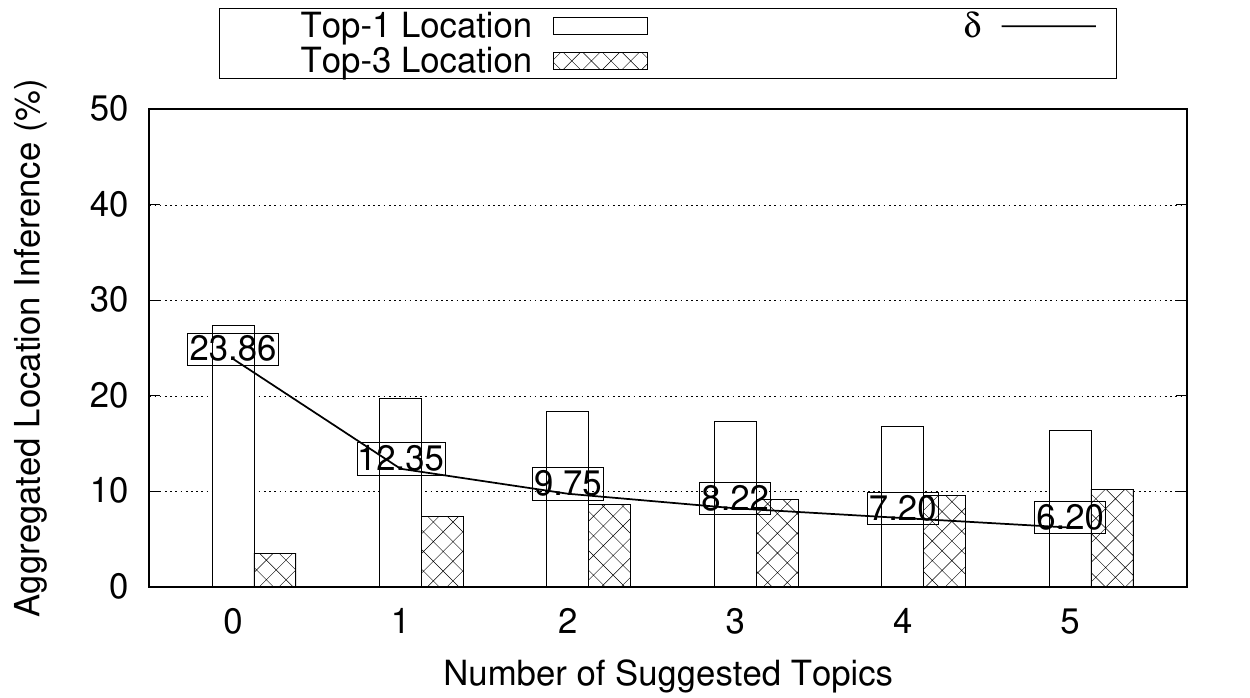}
        \caption{Location (Mild)}
        \label{fig:l_m}
    \end{subfigure}
    \begin{subfigure}[t]{0.32\textwidth}
        \includegraphics[width=\columnwidth]{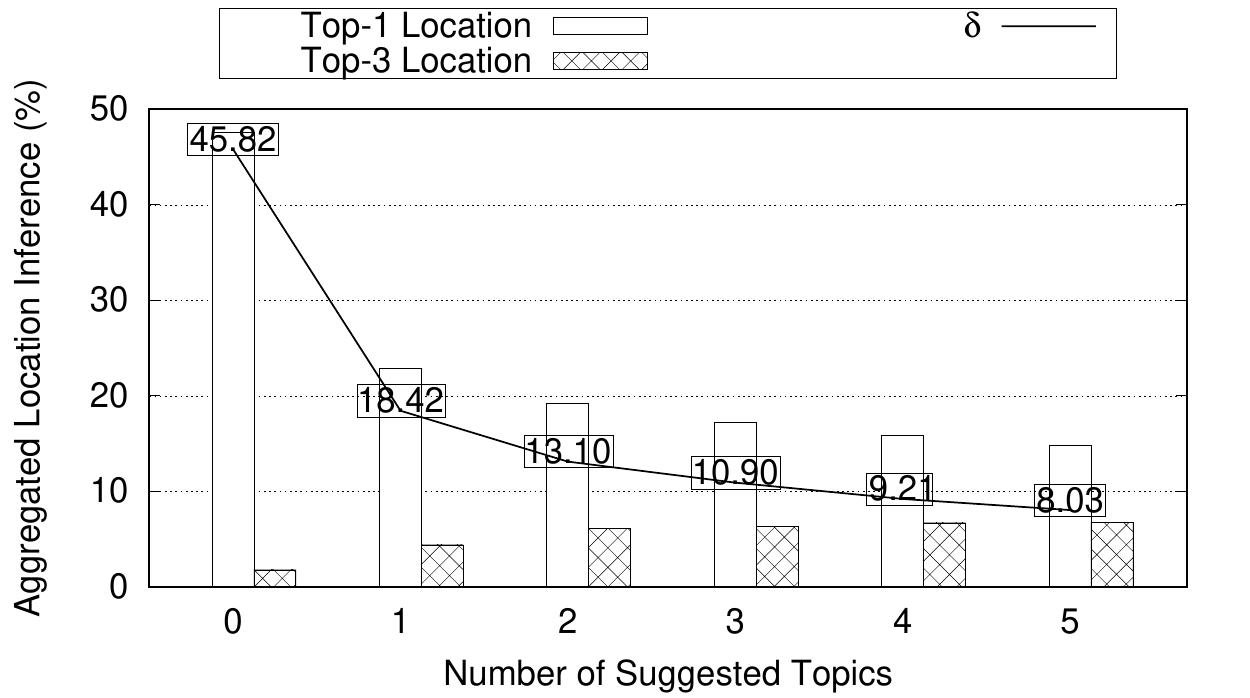}
        \caption{Location (Strong)}
        \label{fig:l_s}
    \end{subfigure}
    
        \begin{subfigure}[t]{0.32\textwidth}
        \includegraphics[width=\columnwidth]{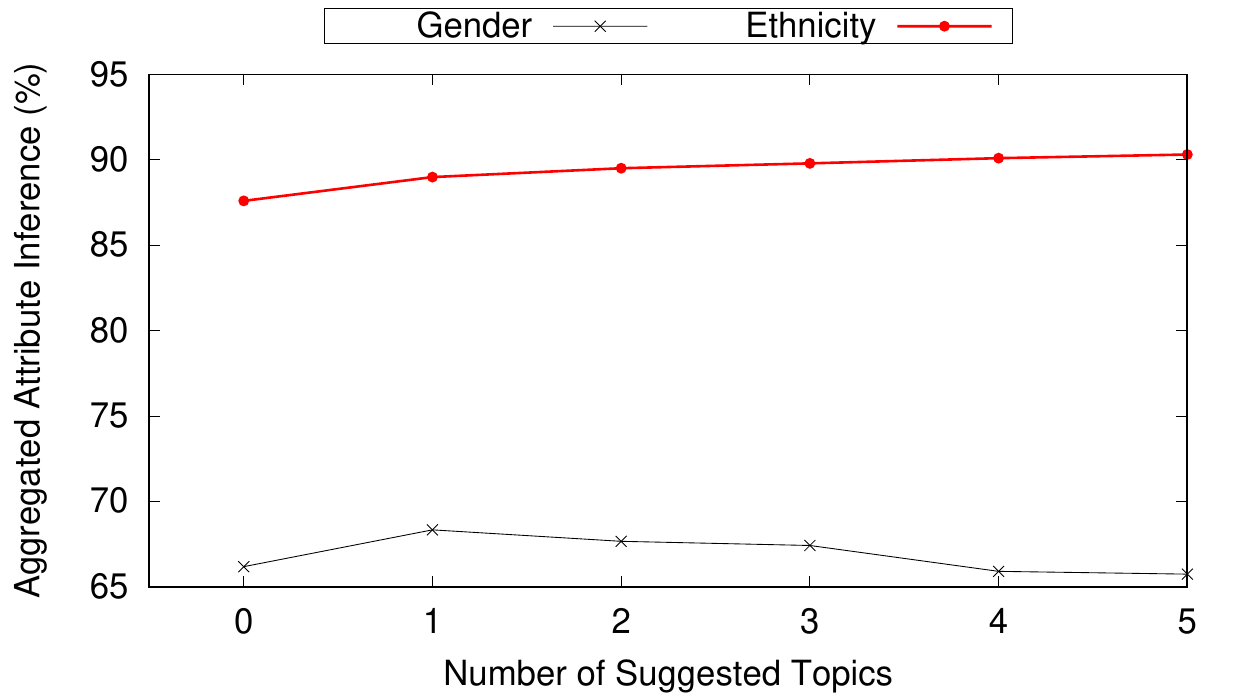}
        \caption{Persona (Weak)}
        \label{fig:l_p_w}
    \end{subfigure}
    \begin{subfigure}[t]{0.32\textwidth}
        \includegraphics[width=\columnwidth]{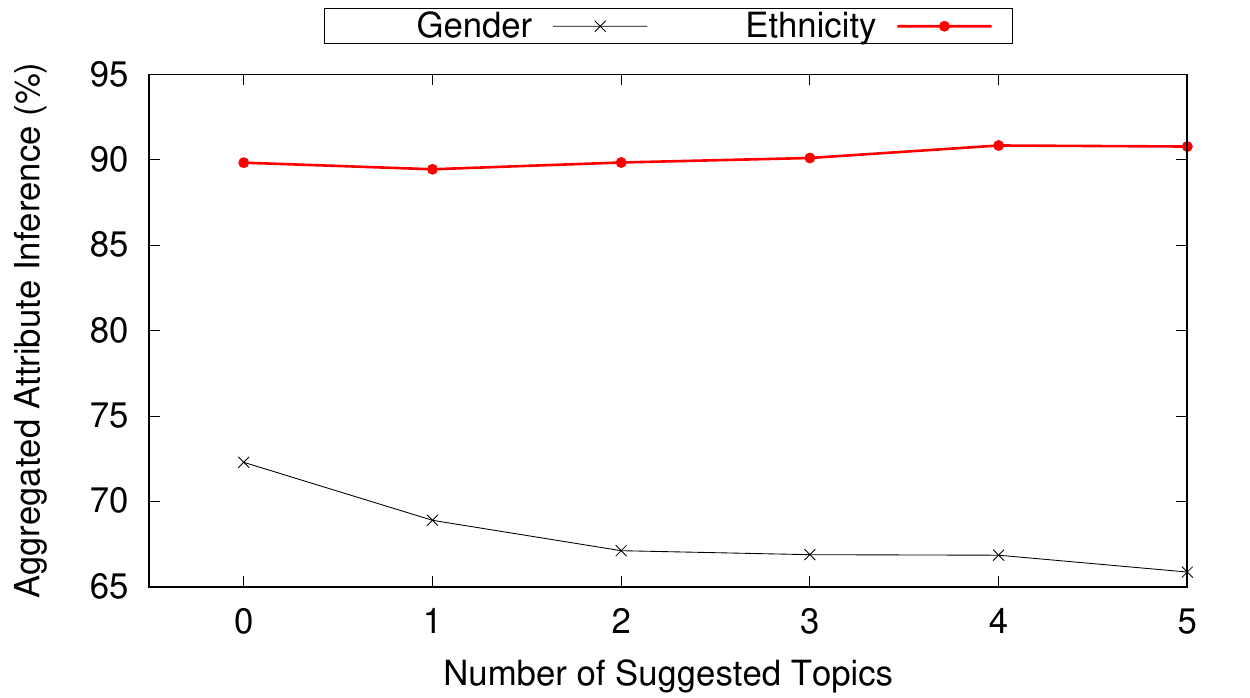}
        \caption{Persona (Mild)}
        \label{fig:l_p_m}
    \end{subfigure}
    \begin{subfigure}[t]{0.32\textwidth}
        \includegraphics[width=\columnwidth]{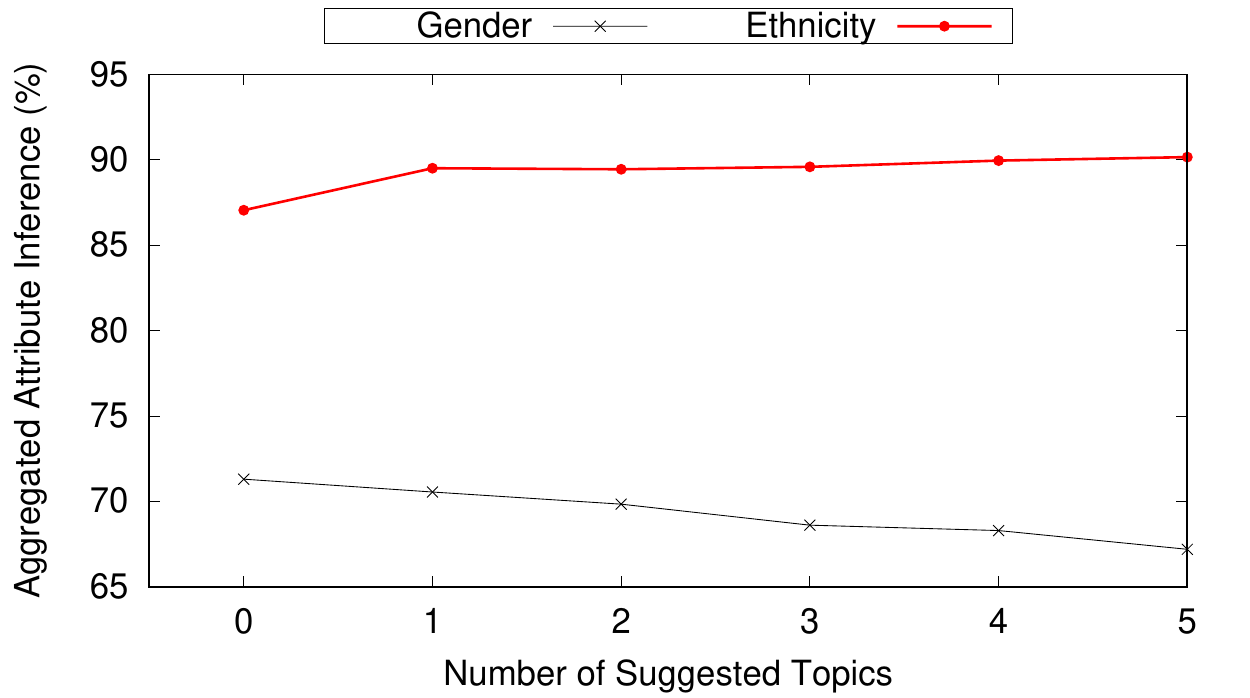}
        \caption{Persona (Strong)}
        \label{fig:l_p_s}
    \end{subfigure} 
    \caption{Effect of obfuscation posts on location and user
    persona given weak, mild and strong connected topics to locations}
    \label{fig:location_obfuscation}
\end{figure*}

In this section, we provide an illustrative obfuscation example that shows
how \sysname achieves multifaceted privacy. This example begins with a newly
created Twitter profile of a Male, White user who lives in California. The user wants to
obfuscate his gender among the gender domain values \{male, female\} achieving
2-gender-indistinguishably. Indistinguishability holds if the privacy
parameter $\Delta_g = 0.1$ is achieved. $\Delta_g$ is set to 0.1 (or 10\%) to ensure
that users who write topics with only negligible topic to persona strength connection
do not have to add any obfuscation posts to their timelines as the topics
they post do not reveal their sensitive attributes. In addition, 
the user wants to preserve his ethnicity and location attributes as his 
public persona. Now, assume that the user tweets \#gowarriors to show his 
support for his favorite Californian basketball team, the Golden State Warriors.

Unfortunately, as shown in Figure~\ref{fig:example-suggestions}, 
\#gowarriors has a strong connection to the male gender attribute 
value. In Figure~\ref{fig:example-suggestions}, the x-axis represents the
posted hashtags one after the other and the y-axis represents the 
\textit{aggregated} gender inferences for both male and female attribute
values over all the posted hashtags. In addition, $\delta$ represents 
the difference of the aggregated gender inference 
between male and female attribute values. As shown, initially, 
$\delta=43\%$ which indicates a strong link between the user's gender and the 
male attribute value.  As $\delta \ge \Delta_g$, this indicates that 
2-gender-indistinguishably is violated.

Therefore, \sysname suggests to post topics that are mainly discussed by 
White people who live in California but linked to the female gender attribute
value. Figure~\ref{fig:example-suggestions} shows the effect of posting subsequent 
topics on the aggregated gender inference. The topic \#womenintech helps to reduce
the aggregate inference difference to 19\%. \#organicfood brings the difference
down to 11\% and \#bodybuilding reduces it to 7\%. Notice that $\delta = 7\%$
achieves $\delta \le \Delta_g$ and hence 2-gender-indistinguishably is achieved.
Notice that the same example holds if the user's true gender is female. The
goal of \sysname is not to invert the gender attribute value but to achieve
inference indistinguishability among $k_g = 2$ different gender attribute values.

\subsection{User Location Obfuscation}\label{sub:location_obfuscation}

The number and the topic to persona connection strength of the obfuscation 
topics largely depend on the topic to persona \textit{connection strength} of 
the original posts. In this experiment, we show how \sysname is used to hide
user location while preserving their gender and ethnicity. In this experiment, 
we set $k_l = 3$ where user location is hidden among three locations. $\Delta_l$ is set to 
0.1 to indicate that 3-location-indistinguishability is achieved if the
difference between the highest (top-1) aggregated location inference and the $3^{rd}$ 
(top-3) aggregated location inference is less than 10\%. The number of obfuscation posts 
needed and their effect on the user persona are reported. This experiment runs using 4707 
weak topics, 1984 mild topics, and 1106 strong topics collected from Twitter over 
several weeks. This experiment assumes a newly created twitter profile simulated with one 
of the personas in Table~\ref{table:topic_analysis}. 
First, a post with some topic to location connection strength (weak, mild, or strong) is added to 
the user profile. Then, we add the suggested obfuscation posts one at a time to the user's timeline. 
After every added obfuscation post, the location and persona inference are reported. The reported numbers
are aggregated and averaged for every topic to location connection strength category.
Figure~\ref{fig:location_obfuscation} shows the effect of adding obfuscation posts on both
the location inference and the persona inference for weak, mild, and strong topics.
As shown in Figure~\ref{fig:l_w}, the 4707 weak topics on average need just 
\textit{one} obfuscation suggestion to achieve 3-location-indistinguishability
when $\Delta_l = 0.1$. Note that adding more obfuscation posts achieves 
3-location-indistinguishability for smaller $\Delta_l$s (e.g., $\Delta_l=0.05$ (4 suggestions), 
$\Delta_l=0.04$ (5 suggestions)). Also, adding one suggestion post achieves $26.5\%$ reduction
in $\delta$. The same experiment is repeated for mild and strong topics
and the results are reported in Figures~\ref{fig:l_m} and~\ref{fig:l_s} respectively.
Notice that strong topics requires \textit{four} suggestions on average to achieve 
3-location-indistinguishability when $\Delta = 0.1$. Also, in 
Figures~\ref{fig:l_s}, adding a single suggestion post
achieves $49.8\%$ reduction in $\delta$.

Figures~\ref{fig:l_p_w},~\ref{fig:l_p_m}, and~\ref{fig:l_p_s} show
the effect on user persona after adding obfuscation posts to weak, mild, and strong 
posts respectively. User persona is represented by gender and ethnicity.
As obfuscation posts are carefully chosen to align with the user persona, we 
observe negligible changes on the average gender and ethnicity inferences after 
adding obfuscation posts.

\subsection{The Effect of Changing $k_l$} \label{sub:k_effect}

\begin{figure}[!htbp]
  \centering
  \includegraphics[width=0.5\columnwidth]{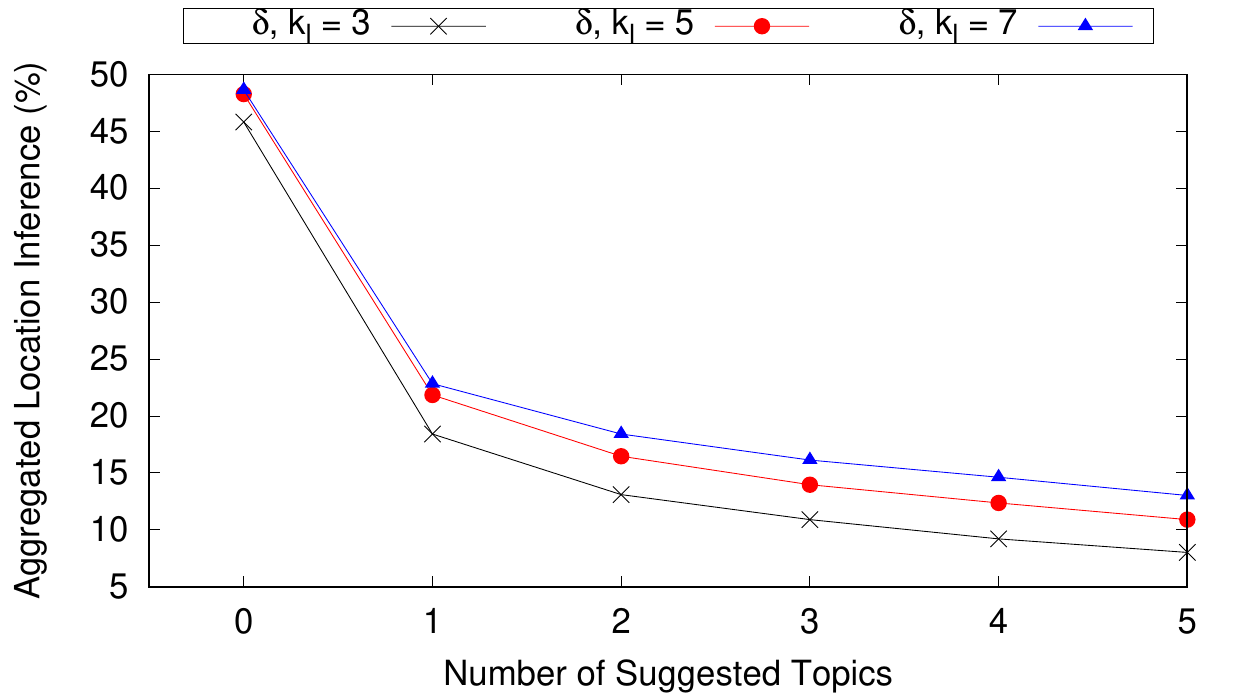}
  \caption{Change in $\delta$ as obfuscation posts are added}
  \label{fig:k_location}
\end{figure}

In this experiment, we measure the effect of changing the parameter $k_l$ on the number 
of obfuscation posts required to achieve k-location-indistinguishability. $k_l$ determines the 
number of locations within which user $u_i$ wants to hide her true location. 
Increasing $k_l$ increases the achieved privacy and boosts the
required obfuscation overhead to achieve k-location-indistinguishability. 
Assume users in Texas hide their State level location among 3 States: 
Texas, Alabama, and Arizona. A malicious advertiser who wants to target users 
in Texas is uncertain about their location and now has to pay 3 times the cost 
of the original advertisement campaign to reach the same target audience. Therefore,
increasing $k$ inflates the cost of micro-targeting.

This experiment uses 621 strong topics collected from Twitter over several weeks.
The average $\delta = V_{u_i}[l]_1 - V_{u_i}[l]_{k_l}$ is reported for all topics. In
addition, the effect of adding obfuscation posts on the average $\delta$ for different values 
of $k_l = 3, 5,$ and $7$ is reported. Figure~\ref{fig:k_location} shows the effect of adding 
obfuscation posts on the aggregated location inference for different values of $k_l$. The privacy
parameter $\Delta_l$ is set to $\Delta_l = 0.1$. 

As shown in Figure~\ref{fig:k_location}, achieving 3-location indistinguishability
for strong topics requires 4 obfuscation posts on the average for
$\Delta_l = 0.1$. On the other hand, 7-location indistinguishability
requires more than 5 obfuscation posts for the same value of $\Delta_l$. 
This result highlights the trade-off between privacy and obfuscation overhead.
Achieving higher privacy levels by increasing $k_l$ or lowering $\Delta_l$ 
requires more obfuscation posts and hence more overhead. Obfuscation posts are
carefully chosen to align with the user persona. Therefore, we observe
negligible changes on the average gender and ethnicity inferences after adding
obfuscation posts for different values of $k_l$.

\subsection{User Gender Obfuscation} \label{sub:gender_obfuscation}
\begin{figure}[!htbp]
  \centering
  \includegraphics[width=0.5\columnwidth]{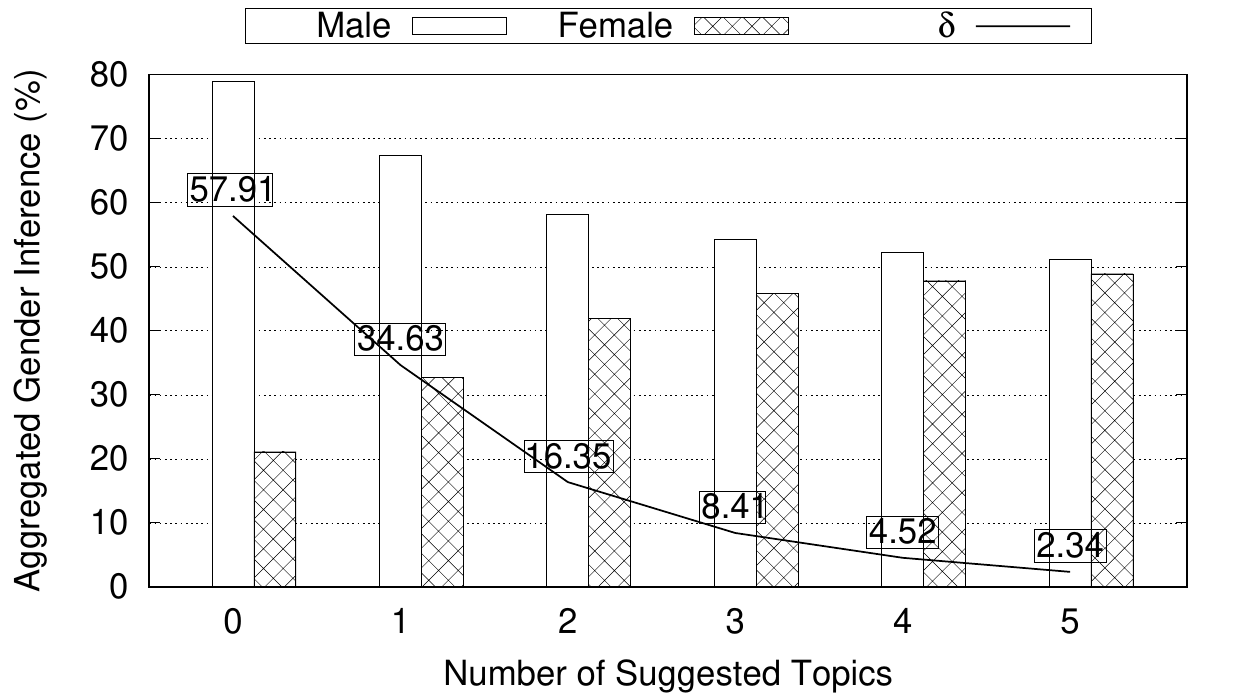}
  \caption{The effect of obfuscation posts on gender inference for strong 
  connected topics to gender.}
  \label{fig:g}
\end{figure}

This experiment shows how \sysname is used to hide
user gender while preserving their ethnicity and location. In this experiment, 
we set $k_g = 2$ where user gender should be hidden among male and female gender domain values. 
$\Delta_g$ is set to $\Delta_g = 0.1$ to indicate that 2-gender-indistinguishability is achieved if 
the difference between the highest (top-1) aggregated gender inference and the $2^{nd}$ 
(top-2) aggregated gender inference is less than 10\%. This experiment runs over 40 gender strongly 
connected topics. The aggregated gender inference is reported when adding the original strong
post to the user's timeline and after adding every obfuscation post one at a time. In addition, 
$\delta$, the difference between the male gender inference and the female gender inference is 
reported. 2-gender-indistinguishability is achieved if $\delta \le 10\%$. As shown in Figure~\ref{fig:g}, 
gender strongly connected topics result in high $\delta$ that violates the 2-gender-indistinguishability 
privacy target. Therefore, \sysname suggest obfuscation posts that results in $\delta$ reduction.
Figure~\ref{fig:g} shows that strong topics need on the average 3 obfuscation posts to achieve
the gender privacy target. This result is quite consistent with the location 
obfuscation experiments in Section~\ref{sub:location_obfuscation}. This shows \sysname's obfuscation
mechanism is quite generic and can be efficiently used to hide different user sensitive attributes.
Finally, as the obfuscation posts are carefully chosen to align with the user's public persona attributes, 
we observed negligible changes on the average location and ethnicity inferences after adding
obfuscation posts.

\section{Related Work}\label{sec:related_work}

The problem of sensitive attribute privacy of social network users has been
extensively studied in the literature from different angles.
\textit{k-anonymity}~\cite{samarati2001protecting,Sweeney:2002,sweeney2002achieving} 
and its successors \textit{l-diversity}~\cite{Machanavajjhala:2007} and 
\textit{t-closeness}~\cite{Li:2007} are well-known and widely used privacy models
in publishing dataset to hide user information among a set of indistinguishable users
in the dataset. Also, differential privacy~\cite{Dwork2006,dwork2008differential,dwork2011differential} has been widely used in 
the context of dataset publishing to hide the identity of a user in a published 
dataset. These models focus on hiding user \textit{identity} 
among other users in the published dataset. Another variation of differential 
privacy is \textit{pan-privacy}~\cite{Dwork:2010-privatestreaming}.
Pan-privacy is designed to work for data streams and hence it is more suitable
for social network streams privacy. However, these models are service 
centric and assume trusted service providers. In this paper, we tackle the privacy
problem from the end-user angle where the user identity is known and all their online
social network postings are public and connected to their identity. Our goal is to
confuse \textit{content-based sensitive attribute inference} attacks by hiding
the user's original public posts among other obfuscation posts. Our multifaceted
privacy achieves the privacy of user sensitive attributes without 
altering their public persona.

In the context of social networks privacy, earlier works focus on sensitive attribute 
inferences due to the structure of social networks. In \cite{zheleva2009join}, Zhelava and Getoor 
attempt to infer the user's sensitive attributes using public and private user profiles. 
However, the authors do not provide a solution to prevent such inference attacks. Georgiou et al. 
~\cite{georgiou2017privacy,georgiou2017extracting} study the inference of sensitive attributes 
in the presence of community-aware trending topic reports. An attacker can 
increase their inference confidence by consuming these reports and the corresponding 
community characteristics of the involved users. In~\cite{georgiou2017privacy}, a mechanism is proposed to prevent social network services from publishing trending
topics that reveal information about individual users. However, this mechanism is
service centric and it is not suitable for hiding a user's sensitive attributes against
content-based inference attacks. Ahmad et al.\cite{ahmad2018intent} introduce a client-centered  
obfuscation solution for protecting user privacy in personalized web searches. The privacy
of a search query is achieved by hiding it among other obfuscation search queries. Although
this work is client-centered, it is not suitable for social networks privacy where the user
online persona has to be preserved.

Recent works have focused on 
the privacy of some sensitive attributes such as location of social network users. Ghufran et 
al.~\cite{ghufran2015toponym} show that social graph analysis can reveal user location
from friends and followers locations. Although, it is important to 
protect user sensitive attributes like location 
against this attack, \sysname focuses only on content-based inference attacks. Yakout et al.~\cite{Yakout:2010} proposed a 
system called Privometer, which measures how much privacy leaks from certain user 
actions (or from their friends' actions) and creates a set of suggestions that could 
reduce the risk of a sensitive attribute being successfully inferred. Similar to Privometer, 
\cite{Heatherly:2013} proposes sanitation techniques to the structure of the social graph by 
introducing noise, and obfuscating edges in the social graphs to prevent 
sensitive information inference. Andres et al.~\cite{Andres:2013} 
propose geo-indistinguishability, a location privacy model that uses differential privacy to hide
a user's exact location in a circle of radius $r$ from
locaction based service providers. In a recent work, Zhang et al.~\cite{zhang2018tagvisor} 
introduce \textit{Tagvisor}, a system to protect users against content-based inference attacks. 
However, Tagvisor requires users to alter their posts by changing or replacing hashtags
that reveal their location. Other works~\cite{gong2015personalized, brown2013haze} depend on user 
collaboration to hide an individual's exact location among the location of the collaboration group.
This approach requires group members to collaborate and synchronously change 
their identities to confuse adversaries. However, these techniques are prone to 
content-based inference attacks and collaboration between users might be hard to 
achieve in the social network context. For location
based services, Mokbel et al.~\cite{mokbel2006new} use 
location generalization and k-anonymity to hide the exact location of a query. These works
do not preserve the user online persona while achieving location privacy. In addition, these works do not provide a generic
mechanism to hide other sensitive attributes such as user gender and ethnicity.

This paper presents \sysname,
the first persona friendly system that 
enables users to hide their sensitive attributes while preserving their online
persona. \sysname is a client-centric solution that can be used to hide any user
specified sensitive attribute. \sysname does not require users to alter their 
original posts or topics. Instead, \sysname hides the user's original posts 
among other obfuscation posts that are aligned with their persona but linked 
to other sensitive attribute values achieving $k$-attribute-indistinguishability.

\section{Future Extensions} \label{sec:future_extensions}

Research in the social network privacy has focused on dataset publishing
and obfuscating user information among other users. Such focus is \textit{service
centric} and assumes that service providers are trusted. However, \sysname is \textit{user
centric} and aims to give the users control over their own privacy. This control comes with
a cost represented by the obfuscation posts that need to be posted by users. The role
of \sysname is to automate the topic suggestion process and to ensure that
k-attribute-indistinguishability holds against content-base inference attacks. There are 
several directions where these obfuscation and privacy models can evolve. 

\textit{Obfuscation Post Generation}. \sysname suggests topics as keywords or hashtags 
and requires users to write the obfuscation posts using their personal writing styles to 
ensure that the original posts and the obfuscation posts are indistinguishable. 
However, the obfuscation writing overhead might
alienate users from \sysname. Instead, \sysname can exploit deep neural network 
language models to learn the user's writing style. 
\sysname can use such a model to suggest full posts instead of hashtags and 
users can either directly publish these posts or edit them before publishing.
This extension aims to reduce the overhead on the users by automating the obfuscation
post generation.

\textit{Social Graph Attack Prevention}. \sysname is mainly designed to 
obfuscate the user sensitive attributes against content-based inference 
attacks. An orthogonal attack is to use the attribute values of friends 
and followers to infer a user's real sensitive attribute value. Take 
location as a sensitive attribute example. Ghufran et 
al.,~\cite{ghufran2015toponym} show that user location can be inferred
from the locations of followers and friends. A user whose friends are 
mostly from NYC is highly probable to be from NYC. \sysname can be 
extended to prevent this attack. Users of similar persona but different 
locations can create an indistinguishability network where users in this
network have followers and friends from different locations. Similar to 
the obfuscation topics, \sysname could suggest users to follow with 
similar persona but different locations.

\section{Conclusion} \label{sec:conclusion}

In this paper, we propose {\em multifaceted privacy}, a novel privacy model 
that obfuscates a user's sensitive attributes while publicly revealing their public online persona. To achieve the multifaceted privacy, we build \textit{\sysname}, a prototype client-centric social network stream processing
system that achieves multifaceted privacy.
\sysname is user-centric and allows social network users to control 
which persona attributes should be publicly revealed and which should be kept 
private. \sysname is designed to transfer user 
trust from the social network providers to her local machine. 
For this, \sysname continuously suggests \textit{topics} and 
\textit{hashtags} to social network users to post in order to obfuscate 
their sensitive attributes and hence confuse content-based sensitive attribute 
inferences. The suggested 
topics are carefully chosen to preserve the user's publicly revealed persona 
attributes while hiding their private sensitive persona attributes.
Our experiments show that \sysname is able to achieve
sensitive attributes privacy such as location and gender. Adding as few as 0 to 4 obfuscation posts (depending on how strongly connected the original post is to a 
persona) successfully hides the user specified sensitive attributes 
without altering
the user's public persona attributes.

\balance
\bibliographystyle{abbrv}
\bibliography{main} 

\end{document}